\documentclass[12pt]{article}
\usepackage{amssymb,amsmath,cite}
\catcode`\@=11

\@addtoreset{equation}{section}
\@addtoreset{equation}{subsection}
\def\theequation{\arabic{section}.\arabic{subsection}.\arabic{equation}}
     
\catcode`\@=11
\def\thesection{\Roman{section}}

\def\appendix{\setcounter{section}{0}
        \def\thesection{Appendix.}
        \def\theequation{\Alph{section}.\arabic{equation}}}
\def\section{\@startsection{section}{1}{\z@}{3.5ex plus 1ex minus
   .2ex}{2.3ex plus .2ex}{\large\bf}}

\long\def\@makefntext#1{\parindent 0cm\noindent
\hbox to 1em{\hss$^{\@thefnmark}$}#1}

\newcommand{\beq}{\begin{equation}}
\newcommand{\eeq}{\end{equation}}
\def\cross{\mathop{{\Bigl\backslash\hspace{-.8em}\Bigr\slash}}}
\def\sidex{\mathop{\Bigl\rangle\Bigr\langle}}
\def\topx{\mathop{\raise4pt\hbox{$\vee$}\hspace{-.7em}\lower4pt\hbox{$\wedge$}}}

\begin{document}
\begin{titlepage}
\vspace{.5in}
\begin{flushright}
UCD-04-24\\
gr-qc/0409039\\
September 2004\\
\end{flushright}
\vspace{.5in}
\begin{center}
{\Large\bf
 Quantum Gravity in 2+1 Dimensions:\\[.6ex]
The Case of a Closed Universe}\\
\vspace{.4in}
{S.~C{\sc arlip}\footnote{\it email: carlip@physics.ucdavis.edu}\\
       {\small\it Department of Physics}\\
       {\small\it University of California}\\
       {\small\it Davis, CA 95616}\\{\small\it USA}}
\end{center}

\vspace{.5in}
\begin{center}
{\large\bf Abstract}
\end{center}
\begin{center}
\begin{minipage}{4.75in}
{\small In three spacetime dimensions, general relativity
drastically simplifies, becoming a ``topological'' theory
with no propagating local degrees of freedom.  Nevertheless,
many of the difficult conceptual problems of quantizing
gravity are still present.  In this review, I summarize the
rather large body of work that has gone towards quantizing 
(2+1)-dimensional vacuum gravity in the setting of a spatially 
closed universe.
}
\end{minipage}
\end{center}
\end{titlepage}
\addtocounter{footnote}{-1}

\section{Introduction}

The task of quantizing general relativity is one of the outstanding 
problems of modern theoretical physics.  Attempts to reconcile 
quantum theory and general relativity date back to the 1930s 
(see \cite{Rovelli} for a historical review), and decades of hard
work have yielded an abundance of insights into quantum field
theory, from the discovery of DeWitt-Faddeev-Popov ghosts to
the development of effective action and background field methods
to the detailed analysis of the quantization of constrained systems.  
But despite this enormous effort, no one has yet succeeded in 
formulating a complete, self-consistent quantum theory of 
gravity \cite{Carliprev}.

The obstacles to quantizing gravity are in part technical.  General 
relativity is a complicated nonlinear theory, and one should expect 
it to be more difficult than, say, electrodynamics.  Moreover, 
viewed as an ordinary field theory, general relativity has a 
coupling constant $G^{1/2}$ with dimensions of an inverse mass, 
and standard power-counting arguments---finally confirmed in
1986 by explicit computations \cite{Goroff}---indicate that the 
theory is nonrenormalizable.  But the problem of finding a consistent 
quantum theory of gravity goes deeper.  General relativity is a 
geometric theory of spacetime, and quantizing gravity means 
quantizing spacetime itself.  In a very basic sense, we do not know 
what this means.  For example:
\begin{itemize}
\item Ordinary quantum field theory is local, but the fundamental 
(diffeomorphism-invariant) physical observables of quantum 
gravity are necessarily nonlocal;
\item Ordinary quantum field theory takes causality as a fundamental 
postulate, but in quantum gravity the spacetime geometry, and thus the
light cones and the causal structure, are themselves subject to quantum
fluctuations;
\item Time evolution in quantum field theory is determined by a
Hamiltonian operator, but for spatially closed universes, the
natural candidate for a Hamiltonian in quantum gravity is identically
zero when acting on physical states;
\item Quantum mechanical probabilities must add up to unity at a
fixed time, but in general relativity there is no preferred time-slicing
on which to normalize probabilities.
\end{itemize}

Faced with such problems, it is natural to look for simpler models that 
share the important conceptual  features of general relativity while 
avoiding some of the computational difficulties.  General relativity in 
2+1 dimensions---two dimensions of space plus one of time---is one
such model.  As a generally covariant theory of spacetime geometry, 
(2+1)-dimensional gravity has the same conceptual foundation as 
realistic (3+1)-dimensional general relativity, and many of the 
fundamental issues of quantum gravity carry over to the lower dimensional 
setting.  At the same time, however, the (2+1)-dimensional model is 
vastly simpler, mathematically and physically, and one can actually 
write down viable candidates for a quantum theory.  With a few exceptions, 
(2+1)-dimensional solutions are physically quite different from those
in 3+1 dimensions, and the (2+1)-dimensional model is not very helpful 
for understanding the dynamics of realistic quantum gravity.  In particular,
the theory does not have a good Newtonian limit \cite{DJtH,Barrow,Cornish}.  But 
for understanding conceptual problems---the nature of time, the construction
of states and observables, the role of topology and topology change, the
relationships among different approaches to quantization---the model 
has proven highly instructive.

Work on (2+1)-dimensional gravity dates back to 1963, when
Staruszkiewicz first described the behavior of static solutions with
point sources \cite{Staruszkiewicz}.  Progress continued sporadically 
over the next twenty years, but the modern rebirth of the subject can 
be traced to the seminal work of Deser, Jackiw, 't~Hooft, and Witten 
in the mid-1980s \cite{DJtH,DesJaca,DesJacb,tHooft,Gerbert,Wittena,Wittenb}.  
Over the past twenty years, (2+1)-dimensional gravity has become 
an active field of research, drawing insights from general relativity, 
differential geometry and topology, high energy particle theory, 
topological field theory, and string theory.   

As I will explain below, general relativity in 2+1 dimensions has
no local dynamical degrees of freedom.  Classical solutions to the vacuum
field equations are all locally diffeomorphic to spacetimes of constant 
curvature, that is, Minkowski, de Sitter, or anti-de Sitter space.  Broadly
speaking, three ways to introduce dynamics have been considered:
\begin{enumerate}
\item Point particles can be added, appearing as conical ``defects'' in 
an otherwise constant curvature spacetime.  Most of the earliest papers
in the field \cite{Staruszkiewicz,DesJaca,DesJacb,DJtH,tHooft,Gerbert} 
were investigations of the dynamics of such conical singularities.
\item If a negative cosmological constant is present, black hole solutions
can be found \cite{BTZ,BHTZ}.  For such solutions, dynamics at either
the horizon or the boundary at infinity can lead to local degrees of
freedom \cite{Carlip1,Strominger,Birmingham,CHvD,BERS,Arc,Chen}, although 
these are certainly not yet completely understood \cite{Carlip2}.
\item One can consider nontrivial spatial or spacetime topologies
\cite{Wittena,Wittenb}.  Such  ``cosmological'' solutions have moduli---%
a finite number of parameters that distinguish among geometrically
inequivalent constant curvature manifolds---and these can become 
dynamical.
\end{enumerate}
In this paper, I will limit myself to the third case, (2+1)-dimensional
vacuum ``quantum cosmology.''  This review is based in part on a series 
of lectures in \cite{CarKorea} and an earlier review \cite{Carsix}, 
and much of the material can be found in more detail in the book
\cite{Carlipbook}.  There is not yet a comprehensive review of  
gravitating point particles in 2+1 dimensions, although Refs.\ 
\cite{Carlip0,Matschull,Matschull1,Bais,Bais2,Menotti,Cantini,Louko} will 
give an overview of some results.  Several good general reviews of the 
(2+1)-dimensional black hole exist \cite{Carlip3,Banados}, although a 
great deal of the quantum mechanics is not yet understood \cite{Carlip2}.

Although string theory is perhaps the most popular current approach to
quantum gravity, I will have little to say about it here: while some
interesting results exist in 2+1 dimensions, almost all of them are in 
the context of black holes (see, for example, \cite{HorWelch,Kal,MalOog1,%
MalOog2,MalOog3}).  I will also have little to say about (2+1)-dimensional 
supergravity, although many of the results described below can be generalized 
fairly easily, and I will not address the coupling of matter except for a 
brief discussion in section \ref{whatb}.

Throughout, I will use units $16\pi G=1$ and $\hbar=1$
unless otherwise noted.

\section{Classical Gravity in 2+1 Dimensions}

The first step towards quantizing (2+1)-dimensional general relativity
is to understand the space of classical solutions.  One of the principal
advantages of working in 2+1 dimensions is that for simple enough
topologies, this space can be characterized completely and explicitly.
Indeed, there are several such characterizations, each leading naturally 
to a different approach to the quantum theory; by understanding 
the relationships among these approaches, one can gain important
insights into the structure of quantum gravity.

\subsection{Why (2+1)-dimensional gravity is simple}

In any spacetime, the curvature tensor
may be decomposed into a curvature scalar $R$, a Ricci tensor
$R_{\mu\nu}$, and a remaining trace-free, conformally invariant
piece, the Weyl tensor $C_{\mu\nu\rho}{}^{\sigma}$.  In 2+1 dimensions,
however, the Weyl tensor vanishes identically, and the full curvature
tensor is determined algebraically by the remaining pieces:
\beq
R_{\mu\nu\rho\sigma} = g_{\mu\rho}R_{\nu\sigma}
+ g_{\nu\sigma}R_{\mu\rho} - g_{\nu\rho}R_{\mu\sigma}
- g_{\mu\sigma}R_{\nu\rho} - \frac{1}{2}
(g_{\mu\rho}g_{\nu\sigma} - g_{\mu\sigma}g_{\nu\rho})R .
\label{a1}
\eeq
This means that any solution of the field equations with a cosmological 
constant $\Lambda$,
\beq
R_{\mu\nu} = 2\Lambda g_{\mu\nu} ,
\label{a2}
\eeq
has constant curvature: the spacetime is locally either flat ($\Lambda=0$),
de Sitter ($\Lambda>0$), or anti-de Sitter ($\Lambda<0$).   Physically, a 
(2+1)-dimensional spacetime has no local degrees of freedom: there are 
no gravitational waves in the classical theory, and no propagating gravitons 
in the quantum theory.

This absence of local degrees of freedom can be verified by a simple counting 
argument \cite{Barrow,Cornish}.  In $n$ dimensions, the phase space of general
relativity is parametrized by a spatial metric at constant time, which 
has $n(n-1)/2$ components, and its conjugate momentum, which adds another 
$n(n-1)/2$ components.  But $n$ of the Einstein field equations are constraints 
rather than dynamical equations, and $n$ more degrees of freedom can be 
eliminated by coordinate choices.  We are thus left with $n(n-1)-2n = n(n-3)$ 
physical degrees of freedom per spacetime point.  In four dimensions, this 
gives the usual four phase space degrees of freedom, two gravitational wave 
polarizations and their conjugate momenta.  If $n=3$, there are no local 
degrees of freedom.  

It is instructive to examine this issue in the weak field approximation
\cite{Birmingham0}.  In any dimension, the vacuum field equations in harmonic
gauge for a nearly flat metric $g_{\mu\nu} = \eta_{\mu\nu} + h_{\mu\nu}$  
take the form
\beq
\Box {\bar h}_{\mu\nu} =  {\mathcal O}(h^2),  \qquad
\partial_\mu{\bar h}^{\mu\nu} =0  
\label{a3}
\eeq
where ${\bar h}_{\mu\nu} = h_{\mu\nu} - \frac{1}{2}\eta_{\mu\nu}
\eta^{\rho\sigma}h_{\rho\sigma}$ and indices are raised and lowered 
with the flat metric $\eta$.  The plane wave solutions of (\ref{a3}) are,
to first order,
\beq
{\bar h}_{\mu\nu} = \epsilon_{\mu\nu}e^{ik\cdot x} \qquad
\hbox{with $k^2=0$ and $k^\mu\epsilon_{\mu\nu}=0$} .
\label{a4}
\eeq
Choosing a second null vector $n^\mu$ with $n\cdot k=-1$ and
a spacelike unit vector $m^\mu$ with $k\cdot m = n\cdot m = 0$, 
we can construct a (2+1)-dimensional analog of the Newman-Penrose
formalism \cite{Dreyer}; the polarization tensor $\epsilon_{\mu\nu}$
then becomes
\beq
\epsilon_{\mu\nu} = Ak_\mu k_\nu + B(k_\mu m_\nu + k_\nu m_\mu)
   + C m_\mu m_\nu ,
\label{a5}
\eeq
apparently giving three propagating polarizations.  There is, however,
a residual symmetry: a diffeomorphism generated by an infinitesimal vector 
field $\xi^\mu$ with $\Box\xi^\mu=0$
preserves the harmonic gauge condition of (\ref{a3}) while giving a 
``gauge transformation'' $\delta{\bar h}_{\mu\nu} = \partial_\mu\xi_\nu 
+ \partial_\nu\xi_\mu  - \eta_{\mu\nu}\partial_\rho\xi^\rho$.  Writing
\beq
\xi_\mu = (\alpha k_\mu + \beta n_\mu + \gamma m_\mu)e^{ik\cdot x} ,
\label{a6}
\eeq
it is easy to check that
\beq
\delta\epsilon_{\mu\nu} = 2i\alpha k_\mu k_\nu 
   + i\gamma(k_\mu m_\nu + k_\nu m_\mu) + i\beta m_\mu m_\nu .
\label{a7}
\eeq
The excitations (\ref{a5}) are thus pure gauge, confirming the absence
of propagating degrees of freedom.
 
Fortunately, while this feature makes the theory simple, it does not quite
make it trivial.  A flat spacetime, for instance, can always be described as
a collection of patches, each isometric to Minkowski space, that are glued
together by isometries of the flat metric; but the gluing is not unique, and
may be dynamical.  This picture leads to the description of (2+1)-dimensional 
gravity in terms of ``geometric structures.''

\subsection{Geometric structures \label{geomstruc}}

The global geometry of vacuum spacetimes in 2+1 dimensions is described
mathematically by the theory of geometric structures \cite{Thur,Canary,Gold3,%
SullThur,Carlip4}.  For simplicity, let us begin with the case of a vanishing 
cosmological constant.  If the spacetime manifold $M$ is topologically 
trivial, then by (\ref{a1}), the vacuum field equations imply that $(M,g)$ 
is simply a subset of ordinary Minkowski space $(V^{2,1},\eta)$.  If $M$ 
is topologically nontrivial, it can still be covered by contractible coordinate 
patches $U_i$, each isometric to $V^{2,1}$, with the standard 
Minkowski metric $\eta_{\mu\nu}$ on each patch.  The geometry is 
then encoded entirely in the transition functions $g_{ij}$ on the 
intersections $U_i\cap U_j$, which determine how these patches are glued 
together.  Since the metrics in $U_i$ and $U_j$ are identical, 
these transition functions must be isometries of $\eta_{\mu\nu}$, that is, 
elements of the Poincar\'e group $\mathrm{ISO}(2,1)$.  Similarly, if 
$\Lambda\ne0$, a vacuum spacetime can be built by patching together
pieces of de Sitter or anti-de Sitter space by appropriate isometries:
$\mathrm{SO}(3,1)$ for $\Lambda>0$ and $\mathrm{SO}(2,2)$ or
$\mathrm{SL}(2,\mathbb{R})\times\mathrm{SL}(2,\mathbb{R})/
\mathbb{Z}_2$ for $\Lambda<0$.

Such a construction is an example of a geometric structure, in the flat 
case a Lorentzian or (ISO(2,1),$V^{2,1}$) structure.  In general, a 
$(G,X)$ manifold is one locally modeled on $X$, much as an ordinary 
$n$-dimensional manifold is modeled on ${\mathbb R}^n$.  More 
precisely, let $G$ be a Lie group that acts analytically on some $n$-manifold 
$X$, the model space, and let $M$ be another $n$-manifold.  A $(G,X)$ 
structure on $M$ is then a set of coordinate patches $U_i$ for $M$ 
with ``coordinates'' $\phi_i: U_i\rightarrow X$ taking their values in  
$X$ and with transition functions $g_{ij}  = \phi_i \circ\phi_j{}^{-1}
|U_i\cap U_j$ in $G$.   

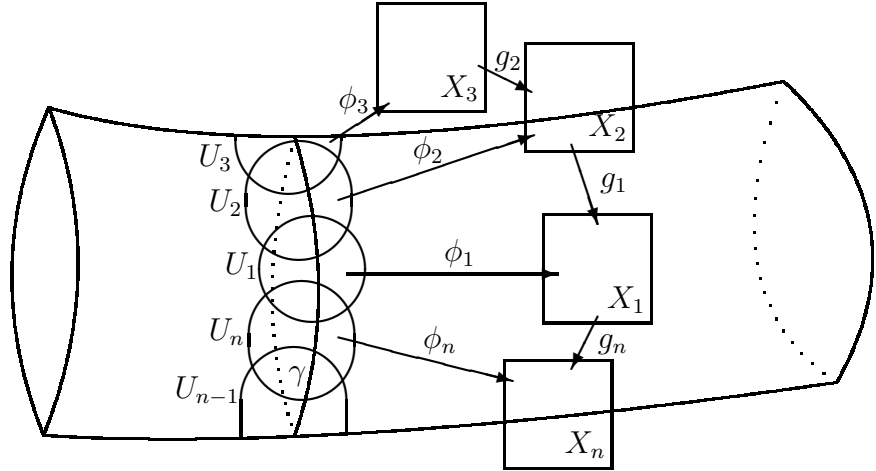
\begin{figure}
\begin{center}
\begin{picture}(300,160)(-70,-30)
\thicklines
\put (38,13){\oval(40,45)}
\put (42,40){\circle{40}}
\put (37,66){\oval(40,45)}
\put (35,-23){\oval(40,67)[t]}
\put (15,-24){\line(0,1){3}}
\put (33,91){\oval(40,45)[b]}
\put (52,66){\vector(3,1){74}}
\put (80,82){$\phi_2$}
\put (55,38){\vector(1,0){80}}
\put (92,43){$\phi_1$}
\put (52,14){\vector(4,-1){66}}
\put (84,10){$\phi_n$}
\put (49,88){\vector(3,2){22}}
\put (52,101){$\phi_3$}
\put (123,85){\framebox(40,40)[br]{\raisebox{1ex}{$X_2\,$}}}
\put (130,20){\framebox(40,40)[br]{\raisebox{1ex}{$X_1\,$}}}
\put (115,-35){\framebox(40,40)[br]{\raisebox{1ex}{$X_n\,$}}}
\put (67,100){\framebox(40,40)[br]{\raisebox{1ex}{$X_3\,$}}}
\put (3,13){$U_n$}
\put (9,38){$U_1$}
\put (3,62){$U_2$}
\put (-1,79){$U_3$}
\put (-10,-9){$U_{n-1}$}
\put (140,87){\vector(1,-3){10}}
\put (151,71){$g_1$}
\put (150,22){\vector(-1,-2){10}}
\put (150,10){$g_n$}
\put (105,117){\vector(2,-1){20}}
\put (111,117){$g_2$}
\qbezier(-60,-23)(50,-30)(240,-3)      
\qbezier(-58,101)(33,75)(220,111)      
\qbezier(35,-23)(53,30)(35,90)         
\qbezier[25](35,-23)(18,27)(35,90)     
\put (33,-2){$\gamma$}
\qbezier(-60,-23)(-35,40)(-58,101)      
\qbezier(-60,-23)(-85,40)(-58,101)      
\qbezier(240,-3)(275,59)(220,111)       
\qbezier[20](240,-3)(190,30)(220,111)   
\end{picture}
\end{center}
\caption{\small The curve $\gamma$ is covered by coordinate patches 
$U_i$, with transition functions $g_i \in G$.  The composition $g_1\circ 
\dots\circ g_n$ is the holonomy of the curve.\label{gstruc}}
\end{figure}

A fundamental ingredient in the description of a $(G,X)$ structure is
its holonomy group, which can be viewed as a measure of the failure of a
single coordinate patch to extend around a closed curve.  Let $M$ be a
$(G,X)$ manifold containing a closed path $\gamma$.  As illustrated in 
figure \ref{gstruc}, we can cover $\gamma$ with coordinate charts
\beq
\phi_i: U_i\rightarrow X,\qquad i=1,\dots,n
\label{ab1}
\eeq
with constant transition functions $g_i\in G$ between $U_i$ and $U_{i+1}$,
i.e.,
\begin{align}
\phi_i|U_i\cap U_{i+1} &= g_i\circ \phi_{i+1}|U_i\cap U_{i+1}\nonumber\\
\phi_n|U_n\cap U_{1} &= g_n\circ \phi_{1}|U_n\cap U_{1} .
\label{ab2}
\end{align}
Let us now try to analytically continue the coordinate $\phi_1$ from the
patch $U_1$ to the whole of $\gamma$.  We can begin with a coordinate
transformation in $U_2$ that replaces $\phi_2$ by ${\phi_2}'=g_1\circ\phi_2$,
thus extending $\phi_1$ to $U_1\cup U_2$.  Continuing this process along
the curve, with ${\phi_j}' = g_1\circ\dots\circ g_{j-1}\circ\phi_j$, we will
eventually reach the final patch $U_n$, which again overlaps $U_1$.  If the
new coordinate function ${\phi_n}'=g_1\circ\dots\circ g_{n-1}\circ\phi_n$
agrees with $\phi_1$ on $U_n\cap U_1$, we will have covered $\gamma$ with 
a single patch.  Otherwise, the holonomy  $H(\gamma) = g_1\circ\dots\circ g_n$ 
measures the obstruction to such a covering.

It may be shown that the holonomy of a curve $\gamma$ depends only on its
homotopy class \cite{Thur}.  In fact, the holonomy defines a homomorphism
\beq
H: \pi_1(M)\rightarrow G .
\label{ab3}
\eeq
$H$ is not quite uniquely determined by the geometric
structure, since we are free to act on the model space $X$ by a fixed
element $h\in G$, changing the transition functions $g_i$ without
altering the $(G,X)$ structure of $M$.  Such a transformation has the 
effect of conjugating $H$ by $h$, and it may be shown that $H$ is  
unique up to such conjugation \cite{Thur}.  The space of holonomies 
is thus the quotient
\begin{align}
&\mathcal{M} = \mathop{Hom}(\pi_1(M), G)/\sim  \nonumber\\
& \rho_1\sim\rho_2 \ \hbox{if $\rho_2 = h\cdot\rho_1\cdot h^{-1}$, $h\in G$} .
\label{ab4}
\end{align}

Note that if we pass from $M$ to its universal covering space $\widetilde
M$, we will no longer have noncontractible closed paths, and $\phi_1$ will
be extendible to all of $\widetilde M$.  The resulting map $D: \widetilde
M\rightarrow X$ is called the developing map.
At least in simple examples, $D$ embodies the classical geometric
picture of development as ``unrolling''---for instance, the unwrapping
of a cylinder into an infinite strip.

The holonomies of the geometric structure in (2+1)-dimensional gravity
are examples of diffeomorphism-invariant observables, which, as we shall 
see below, are closely related to the Wilson loop observables in the
Chern-Simons formulation.  It is important to understand
to what extent they are complete---that is, to what extent they determine
the geometry.  It is easy to see one thing that can go wrong: if we start with 
a flat three-manifold $M$ and simply cut out a ball, we can obtain a new
flat manifold without affecting the holonomy.  This is a rather trivial 
change, though, and we would like to know whether it is the only problem.

For the case of a vanishing cosmological constant, Mess \cite{Mess} has 
investigated this question for spacetimes with topologies  
$\mathbb{R}\times\Sigma$.  He shows that the holonomy group 
determines a unique ``maximal'' spacetime $M$---specifically, a  
domain of dependence of a spacelike surface 
$\Sigma$.  Mess also demonstrates that the holonomy group $H$ acts 
properly discontinuously on a region $W\!\subset\!V^{2,1}$ of Minkowski 
space, and that $M$ can be obtained as the quotient space $W/H$.  This 
quotient construction can be a powerful tool for obtaining a description of 
$M$ in reasonably standard coordinates, for instance in a time-slicing by 
surfaces of constant mean curvature.  Similar results hold for anti-de
Sitter structures.  Some instructive examples of the construction of 
spacetimes with $\Lambda<0$ from holonomies are given in \cite{Fujiwara}.

For de Sitter structures, on the other hand, the holonomies do {\it not\/}
uniquely determine the geometry \cite{Mess}.  An explicit example of
the resulting ambiguity has been given by Ezawa \cite{Ezawa} for the
case of a topology $\mathbb{R}\times T^2$ (see also section 4.5 of 
\cite{Carlipbook}).  A similar ambiguity occurs for (2+1)-dimensional
gravity with point particles, where, as Matschull has emphasized
\cite{Matschull2}, it may imply a physical difference between the metric
and Chern-Simons formulations of (2+1)-dimensional gravity.

We close this section with a partial description of the space of solutions 
of the vacuum Einstein field equations on a manifold $\mathbb{R}\times 
\Sigma$, where $\Sigma$ is a compact genus $g$ two-manifold, that
is, a surface with $g$ ``handles.''  The fundamental group of such a 
spacetime, $\pi_1(M)\simeq\pi_1(\Sigma)$, is generated by $g$ pairs
of closed curves $(A_i,B_i)$, with the single relation
\beq
A_1B_1A_1{}^{-1}B_1{}^{-1}A_2B_2A_2{}^{-1}B_2{}^{-1}\dots
A_gB_gA_g{}^{-1}B_g{}^{-1} = 1 .
\label{ab5}
\eeq
By (\ref{ab4}), the space of holonomies is the 
space of homomorphisms from $\pi_1(\Sigma)$ to $G$ (where
$G$ is $\mathrm{ISO}(2,1)$ for $\Lambda=0$, $\mathrm{SO}(3,1)$
for $\Lambda>0$, or $\mathrm{SO}(2,2)$ for $\Lambda<0$) modulo
overall conjugation.  For $g>1$, this space  of homomorphisms has 
dimension $12g-12$: $\pi_1(\Sigma)$ has $2g$ generators and one 
relation, and the identification by conjugation leaves $2g-2$ choices 
of elements of a six-dimensional group $G$.\footnote{For $g=0$,
$\pi_1(\Sigma)$ is trivial, and there is only one geometric structure.
The case of $g=1$ will be discussed below in section \ref{tor}.}  

There are two subtleties that prevent the space (\ref{ab4}) from being
the exact moduli space of solutions of the vacuum field equations.
First, as noted above, the holonomies do not always determine a
unique geometric structure.  In particular, for $\Lambda>0$ one
may need an additional discrete variable to specify the geometry.
Second, not all homomorphisms from $\pi_1(\Sigma)$ to $G$ give 
geometric structures that correspond to smooth manifolds.  
The space of homomorphisms (\ref{ab4}) is not connected \cite{Gold1}, 
and, in general, only one connected component gives our desired
geometry.  Even once these caveats are taken into account, though, 
we still have a $(12g-12)$-dimensional space of solutions that can,
in principle, be described completely.
 
\subsection{The Chern-Simons formulation \label{first}}

The formalism of geometric structures provides an elegant description 
of vacuum spacetimes in 2+1 dimensions, but it is rather remote from
the usual physicist's approach.  In particular, the  Einstein-Hilbert action
is nowhere in sight, and even the metric makes only a limited appearance.
Fortunately, the description is closely related to the more familiar first-order
Chern-Simons formalism \cite{Deser,Wittena,Wittenb,Achucarro}, which, in 
turn, can connect us back to the standard metric formalism.

The first-order formalism takes as its fundamental variables an
orthonormal frame (``triad'' or ``dreibein'') $e_\mu{}^a$, which
determines a metric $g_{\mu\nu} = \eta_{ab}e_\mu{}^a e_\nu{}^b$,
and a spin connection $\omega_\mu{}^{ab}$.  As in the Palatini formalism, 
$e$ and $\omega$ are treated as independent quantities.  In terms of
the one-forms
\beq
e^a = e_\mu{}^a dx^\mu , \qquad 
\omega^a = \frac{1}{2}\epsilon^{abc} \omega_{\mu bc}dx^\mu ,
\label{ac1}
\eeq
the first-order action takes the form
\beq
I = 2\int_M \left\{ e^a\wedge\left( d\omega_a 
  + \frac{1}{2}\epsilon_{abc} \omega^b\wedge \omega^c \right)
  + \frac{\Lambda}{6}\epsilon_{abc} e^a\wedge e^b\wedge e^c \right\} ,
\label{ac2}
\eeq
with Euler-Lagrange equations  
\begin{align}
T_a &= de_a + \epsilon_{abc}\omega^b\wedge e^c = 0 ,
\label{ac3}\\
R_a &= d\omega_a + \frac{1}{2}\epsilon_{abc}\omega^b\wedge\omega^c 
  = -\frac{\Lambda}{2}\epsilon_{abc} e^b\wedge e^c .
\label{ac4}
\end{align}
The first of these implies that the connection is torsion-free, and, if $e$
is invertible, that $\omega$ has the standard expression in terms of  the
triad.  Given such a spin connection, equation (\ref{ac4}) is then equivalent 
to the standard Einstein field equations.  

The action (\ref{ac2}) has two sets of invariances, the local Lorentz 
transformations
\beq
\delta e^a = \epsilon^{abc}e_b\tau_c ,\quad
\delta\omega^a = d\tau^a + \epsilon^{abc}\omega_b\tau_c ,
\label{ac5}
\eeq
and the ``local translations''
\beq
\delta e^a = d\rho^a + \epsilon^{abc}\omega_b\rho_c ,\quad
\delta\omega^a = -\Lambda\epsilon^{abc}e_b\rho_c.
\label{ac6}
\eeq
Provided the triad $e$ is invertible, the latter are equivalent to diffeomorphisms 
on shell; more precisely, the combination of transformations with parameters 
$\rho^a = \xi\cdot e^a$ and $\tau^a = \xi\cdot\omega^a$ is equivalent to the 
diffeomorphism generated by the vector field $\xi$.  The invertibility condition
for $e$ is important; if it is dropped, the first-order formalism is no longer
quite equivalent to the metric formalism \cite{Matschull2}.

As first noted by Ach{\'u}carro and Townsend \cite{Achucarro} and further
developed by Witten \cite{Wittena,Wittenb}, the first-order action (\ref{ac2})
is equivalent to that of a Chern-Simons theory.  Consider first the
case of a vanishing cosmological constant.  The relevant gauge group---the 
group $G$ of the geometric structure---is 
then the Poincar\'e group $\mathrm{ISO}(2,1)$, with standard generators 
${\mathcal J}^a$ and ${\mathcal P}^a$ and commutation relations
\beq
\left[ {\cal J}^a, {\cal J}^b \right] = \epsilon^{abc}{\cal J}_c ,\quad
\left[ {\cal J}^a, {\cal P}^b \right] = \epsilon^{abc}{\cal P}_c ,\quad
\left[ {\cal P}^a, {\cal P}^b \right] = 0  .
\label{ac7}
\eeq
The corresponding gauge potential is
\beq
A = e^a{\cal P}_a + \omega^a{\cal J}_a .
\label{ac8}
\eeq
If one defines a bilinear form (or ``trace'')
\beq
\mathrm{Tr}({\cal J}^a{\cal P}^b) = \eta^{ab} , \qquad
\mathrm{Tr}({\cal J}^a{\cal J}^b) = \mathrm{Tr}({\cal P}^a{\cal P}^b) = 0 ,
\label{ac9}
\eeq
it is straightforward to show that the action (\ref{ac2}) can be
written as
\beq
I_{\mathit{CS}}[A] = \frac{k}{4\pi} \int_M \mathrm{Tr}
  \left\{ A\wedge dA + \frac{2}{3} A\wedge A\wedge A \right\} 
\label{ac10}
\eeq
with $k=1/4G$.  Equation (\ref{ac10}) may be recognized as the standard
Chern-Simons action \cite{Wittenc} for the group $\mathrm{ISO}(2,1)$.

A similar construction is possible when $\Lambda\ne0$.  For $\Lambda 
= -1/\ell^2<0$, the pair of one-forms $A^{(\pm)a} = \omega^a \pm e^a/\ell$
together constitute an $\mathrm{SO}(2,1)\times\mathrm{SO}(2,1)$ gauge 
potential, with a Chern--Simons action 
\beq
I[A^{(+)},A^{(-)}] = I_{\mathit{CS}}[A^{(+)}] 
  - I_{\mathit{CS}}[A^{(-)}]
\label{ac11}
\eeq
that is again equivalent to (\ref{ac2}), provided we set $k= {\ell/4G}$.  
If $\Lambda>0$, the complex one-form $A^a = \omega^a + i\sqrt{\Lambda}e^a$ 
may be viewed as an $\mathrm{SL}(2,\mathbb{C})$ gauge potential, 
whose Chern-Simons action is again equivalent to the first-order 
gravitational action.  For any value of $\Lambda$, it is easily checked
that the transformations (\ref{ac5}) are just the gauge transformations
of $A$.  Vacuum general relativity in 2+1 dimensions is thus equivalent---again 
up to considerations of the invertibility of $e$---to a gauge theory.

We can now connect the first-order formalism to the earlier description of 
geometric structures.  The field equations coming from the action (\ref{ac10}) 
are simply
\beq
F[A] = dA + A\wedge A = 0 ,
\label{ac12}
\end{equation}
implying that the field strength of the gauge potential $A$ vanishes, i.e.,
that $A$ is a flat connection.  Such a connection is
completely determined by its holonomies, that is, by the Wilson loops
\beq
U_\gamma = P\exp\left\{ -\int_\gamma A \right\}
\label{ac13}
\eeq
around closed noncontractible curves $\gamma$.  This use of the term
``holonomy'' is somewhat different from that of section \ref{geomstruc},
but the two are equivalent.  Indeed, any $(G,X)$ structure on 
a manifold $M$ determines a corresponding flat $G$ bundle \cite{Gold3}:
we simply form the product $G\times U_i$ in each patch, giving the local 
structure of a $G$ bundle, and use the transition functions $g_{ij}$ of 
the geometric structure to glue the fibers on the overlaps.  The
holonomy group of this flat bundle can be shown to be isomorphic to the 
holonomy group of the geometric structure, and for (2+1)-dimensional
gravity, the flat connection constructed from the geometric structure is 
that of the Chern-Simons theory.  An explicit construction may be found
in section 4.6 of \cite{Carlipbook}; see also \cite{Amanoa,Unruh}.

The first-order action allows us an additional step that was  
unavailable in the geometric structure formalism---we can compute the
symplectic structure on the space of solutions.  The basic Poisson brackets
follow immediately from the action:
\beq
\{ e_i{}^a(x), \omega_j{}^b(x') \}
  = \frac{1}{2} \eta^{ab}\epsilon_{ij}\delta^2(x-x') .
\label{ac14}
\eeq
The resulting brackets among the holonomies have been evaluated by
Nelson, Regge, and Zertuche \cite{NelReg,NelRegZer} for  
$\Lambda<0$, for which the two $\mathrm{SL}(2,\mathbb{R})$ 
factors in the gauge group $G$ may be taken to be independent.  The
brackets are nonzero only for holonomies of curves that intersect,
and can be written in terms of holonomies of ``rerouted'' curves;
symbolically,  
\beq
\Bigl\{\hspace{.1em} \cross\hspace{-.75em}{\lower7pt\hbox{,}}
  \hspace{.4em} \Bigr\} = 
  \pm\frac{1}{4\ell}\epsilon(p) \Bigl(\hspace{.2em} \cross - 2 \sidex
  \hspace{.2em} \Bigr) 
\label{ac15}
\eeq
where $\epsilon(p)$ is the oriented intersection number at the point
$p$ that the curves cross.  The composition of loops implicit 
in the brackets (\ref{ac15}) makes it difficult to find small closed 
subalgebras of the sort needed for quantization.  However, Nelson 
and Regge have succeeded in constructing a small but complete 
(actually overcomplete) set of holonomies on a surface of arbitrary 
genus that form a closed algebra \cite{Nelsonc,Nelsond}, and Loll
has found a complete set of ``configuration space'' variables \cite{Loll}.

By generalizing a discrete combinatorial approach to Chern-Simons 
theory due to Fock and Rosly \cite{Fock} and Alekseev et al.\ 
\cite{Alekseev1,Alekseev2,Alekseev3}, several authors have further 
explored the quantum group structure of these brackets, which can
be expressed in terms of the quantum double of the Lorentz group 
\cite{Bais,Bais2,Buff,Meusburger}.   It is also interesting that 
the symplectic structure obtained in this way is closely related to the 
symplectic structure on the abstract space of loops on $\Sigma$ first 
discovered by Goldman \cite{Goldman4,GoldHam}.

\subsection{The ADM approach \label{adm}}

We next turn to a more traditional approach to classical general 
relativity, the conventional metric formalism
in the space/time splitting of Arnowitt, Deser, and Misner \cite{ADM}.
As Moncrief \cite{Mon} and Hosoya and Nakao \cite{HosNak} have 
shown, this metric formalism can also be used to give a full description 
of the solutions of the vacuum field equations, at least for spacetimes
with the topology $\mathbb{R}\times\Sigma$.

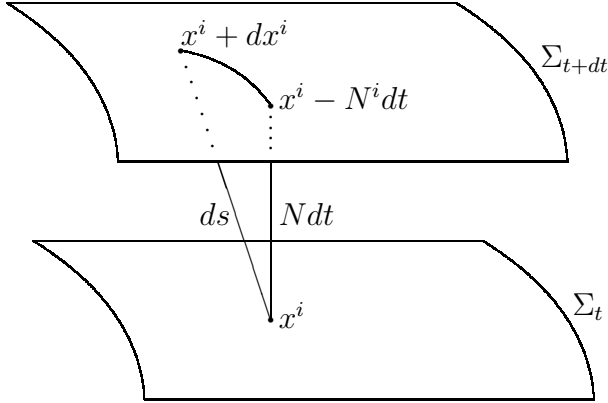
\begin{figure}
\begin{picture}(200,150)(-60,-40)

\qbezier(-0,110)(40,85)(42,50)         
\qbezier(170,110)(210,85)(212,50)
\put(0,110){\line(1,0){170}}
\put(42,50){\line(1,0){170}}
\put(202,86){$\Sigma_{t+dt}$}

\qbezier(10,20)(50,-5)(52,-40)        
\qbezier(180,20)(220,-5)(222,-40)
\put(10,20){\line(1,0){170}}
\put(52,-40){\line(1,0){170}}
\put(214,-8){$\Sigma_t$}

\put(100,-10){\circle*{2}}             
\put(103,-13){$x^i$}

\put(100,-10){\line(0,1){60}}          
\put(100,55){\circle*{1}}
\put(100,59){\circle*{1}}
\put(100,63){\circle*{1}}
\put(100,67){\circle*{1}}
\put(103,26){$Ndt$}

\put(100,71){\circle*{2}}              
\put(103,70){$x^i-N^idt$}

\put(100,-10){\line(-1,3){20}}         
\put(78,56){\circle*{1}}
\put(76,62){\circle*{1}}
\put(74,68){\circle*{1}}
\put(72,74){\circle*{1}}
\put(70,80){\circle*{1}}
\put(68,86){\circle*{1}}
\put(73,26){$ds$}

\put(66,92){\circle*{2}}              
\put(66,94){$x^i+dx^i$}

\qbezier(100,71)(88,88)(66,92)
\end{picture}
\caption{\small The ADM decomposition is based on the Lorentzian 
 version of the Pythagoras theorem. \label{fig2}}
\end{figure}

We start with the ADM decomposition of the spacetime metric 
$g_{\mu\nu}$,
\beq
ds^2 = N^2dt^2 - g_{ij}(dx^i + N^i dt)(dx^j + N^j dt) ,
\label{ad1}
\eeq
as illustrated in figure \ref{fig2}.  The action then takes the usual 
form\footnote{In this section I use standard ADM notation: 
$g_{ij}$ and $R$ refer to the induced metric and scalar curvature 
of a time slice and $K^{ij}$ is the extrinsic curvature of such a slice, 
while the full spacetime metric and curvature are denoted by
${}^{\scriptscriptstyle(3)}\!g_{\mu\nu}$ and 
${}^{\scriptscriptstyle(3)}\!R$.}
\beq
I_{\hbox{\scriptsize grav}}
  =  \int d^3x \sqrt{-{}^{\scriptscriptstyle(3)}\!g}\, 
  ({}^{\scriptscriptstyle(3)}\!R - 2\Lambda)
  = \int dt\int\nolimits_\Sigma d^2x \bigl(\pi^{ij}{\dot g}_{ij}
               - N^i{\cal H}_i -N{\cal H}\bigr) ,
\label{ad2}
\eeq
with canonical momentum $\pi^{ij} = \sqrt{g}\,(K^{ij}- g^{ij}K)$
and the momentum and Hamiltonian constraints 
\beq
{\cal H}_i = -2\nabla_j\pi^j{}_ i,  \qquad
{\cal H} = \frac{1}{\sqrt{g}} g_{ij}g_{kl}(\pi^{ik}\pi^{jl}-\pi^{ij}\pi^{kl})
                 - \sqrt{g}(R - 2\Lambda) .
\label{ad3}
\eeq

To solve the constraints, we can choose the York time-slicing \cite{York},
in which the mean (extrinsic) curvature is used as a time coordinate,
$-K= g_{ij}\pi^{ij}/\sqrt{g}=T$.  Andersson et al.\  have shown that this 
is a good global coordinate choice for classical solutions of the vacuum 
field equations \cite{Andersson}.  We next select a useful parametrization 
of the spatial metric and momentum.  Up to a diffeomorphism, any two-metric 
on $\Sigma$ can be written in the form \cite{Abikoff,Fish}
\beq
g_{ij} = e^{2\lambda}{\bar g}_{ij}(m_\alpha) ,
\label{ad4}
\eeq
where $\bar g_{ij}(m_\alpha)$ are a finite-dimensional family of
metrics of constant curvature $k$ ($k=1$ for the two-sphere,
$0$ for the torus, and $-1$ for spaces of genus $g>1$).  These standard
metrics are labeled by a set of moduli $m_\alpha$ that parametrize 
the Riemann moduli space of $\Sigma$.  As in section \ref{geomstruc}, 
such constant curvature metrics can be described in terms of a geometric 
structure---for genus $g>1$ an  $(\mathbb{H}^2,\mathrm{PSL}(2,\mathbb{R}))$ 
structure---with moduli parametrizing the homomorphisms (\ref{ab3}).  
We can count these just as in section \ref{geomstruc}; now, 
since $\mathrm{PSL}(2,\mathbb{R})$ is three-dimensional, we find that
a constant negative curvature surface of genus $g>1$ is described by 
$6g-6$ parameters.

The corresponding decomposition of the conjugate momentum is 
described in \cite{Mon}: up to a diffeomorphism, the trace-free part 
of $\pi^{ij}$ can be written as a holomorphic quadratic differential
$p^{ij}$, that is, a transverse traceless tensor with respect to the 
covariant derivative compatible with $\bar g_{ij}$.  
The space of such quadratic differentials parametrizes the cotangent 
space of the moduli space \cite{Abikoff}, and the reduced phase 
space becomes, essentially, the cotangent bundle of the moduli
space.

With the decomposition of \cite{Mon}, the momentum constraints 
${\cal H}_i=0$ become trivial, while the Hamiltonian constraint  
becomes an elliptic differential equation that determines the scale 
factor $\lambda$ in (\ref{ad4}) as a function of $\bar g_{ij}$ and 
$p^{ij}$,
\beq
\bar\Delta\lambda - \frac{1}{4}(T^2-4\Lambda)e^{2\lambda}  
  +  \frac{1}{2}\left[  {\bar g}^{-1}
  {\bar g_{ij}(m_\alpha)\bar g_{kl}(m_\alpha)
  p^{ik}(p^\alpha)p^{jl}(p^\alpha)}\right] e^{-2\lambda} - \frac{k}{2} = 0 ,
\label{ad4a}
\eeq
where $p^\alpha$ are the momenta conjugate to the moduli,  
\beq
p^\alpha = \int_\Sigma d^2x\,
  p^{ij}\frac{\partial\ \ }{\partial m_\alpha}{\bar g}_{ij} .
\label{ad6}
\eeq
The theory of elliptic equations ensures that (\ref{ad4a}) determines a
unique scale factor $\lambda$.  The action (\ref{ad2}) then simplifies to
a ``reduced phase space'' action, involving only the physical degrees
of freedom,  
\beq
I_{\hbox{\scriptsize grav}}
  = \int dT \left( p^\alpha \frac{dm_\alpha}{dT} - H(m,p,T) \right) ,
\label{ad5}
\eeq
with a time-dependent Hamiltonian  
\beq
H = \int_{\Sigma_T} d^2x \sqrt{\bar g}\, e^{2\lambda(m,p,T)} .
\label{ad5a}
\eeq
The classical Poisson brackets can be read off directly from (\ref{ad5}):
\beq
\{m_\alpha, p^\beta\} = \delta_\alpha^\beta , \quad
\{m_\alpha, m_\beta\} = \{p^\alpha, p^\beta\} = 0 .
\label{ad7}
\eeq

Three-dimensional gravity again reduces to a finite-dimensional 
system, albeit one with a complicated time-dependent Hamiltonian.  The 
physical phase space is parametrized by $(m_\alpha,p^\beta)$, which may 
be viewed as coordinates for the cotangent bundle of the moduli space of 
$\Sigma$.  For a surface of genus $g>1$, this gives us $12g-12$ degrees of 
freedom, matching the results of section \ref{geomstruc}.  

If $\Lambda=0$, this correspondence can be made more  explicit: for
$G=\mathrm{ISO}(2,1)$ and $M\simeq\mathbb{R}\times\Sigma$, 
the space (\ref{ab4}) of geometric structures is itself a cotangent bundle, 
whose base space is the space of hyperbolic structures on $\Sigma$.  This 
follows from the fact that the group $\mathrm{ISO}(2,1)$ is the cotangent 
bundle of  $\mathrm{SO}(2,1)$.  Concretely, in the first-order formalism
of section \ref{first}, the curvature equation (\ref{ac4}) with $\Lambda=0$
implies that $\omega$ is a flat $\mathrm{SO}(2,1)$ connection; and if 
$\omega(s)$ is a curve in the space of such flat connections, the tangent 
vector $e=d\omega(s)/ds$ satisfies the torsion equation (\ref{ac5}).  
For $\Lambda\ne0$, I know of no such direct correspondence, and the 
general relationship between the ADM and first-order solutions seems 
less transparent.

\subsection{Exact discrete approaches \label{Eda}}

Discrete approximations to general relativity have existed for decades.
In 2+1 dimensions, though, one has the added feature that a discrete
description can be {\it exact}.  This follows from the peculiar nature
of the field equations in three dimensions: as discussed above, any
vacuum solution can be patched together from finite pieces of constant
curvature spacetime, and the dynamics occurs only in the patching.

The ``standard'' discrete approach to classical general relativity is 
Regge calculus \cite{Regge0}, initially developed for (3+1)-dimensional
gravity but extendible to arbitrary dimensions.  Classical Regge
calculus in 2+1 dimensions was investigated by Ro{\v c}ek and Williams 
\cite{Rocek}, who showed that it gave exact results for point particle
scattering.  Regge calculus will be discussed further in section \ref{PR}.

The first discrete formulation designed explicitly for 2+1 dimensions was 
developed by 't Hooft et al.\ \cite{tHooftb,tHooftc,tHooftd,Franzosi,Welling,Holl}.  
This approach has been used mainly to understand point particle 
dynamics, but recent progress has allowed a general
description of topologically nontrivial compact spaces \cite{Kadar}.  
't Hooft's Hamiltonian lattice model is based on the metric formalism, and
starts with a piecewise flat Cauchy surface tessellated by flat polygons,
each carrying an associated frame.  The Einstein field equations with 
$\Lambda=0$ then imply that edges of polygons move at
constant velocities and that edge lengths may change, subject to a
set of consistency conditions.  One obtains a dynamical description
parametrized by a set of lengths and rapidities, which turn out 
to be canonically conjugate.  Complications occur when an edge shrinks 
to zero length or collides with a vertex, but these are completely understood.  
The resulting structure can be simulated on a computer, providing a powerful 
method for visualizing classical evolution in 2+1 dimensions.

A related first-order Hamiltonian lattice model has been studied by 
Waelbroeck et al.\ \cite{Waela,Waelb,Waelc,Waeld}.  This model 
is a discretized version of the first-order formalism of section \ref{first},
with triads assigned to faces of a two-dimensional lattice and Lorentz
transformations assigned to edges.  The model has an extensive gauge 
freedom available in the choice of lattice.  In particular, for a spacetime 
$M\simeq\mathbb{R}\times\Sigma$, one can choose a lattice that 
is simply a $4g$-sided polygon with edges identified; the resulting
spacetime can be visualized as a polygonal tube cut out of Minkowski
spacetime, with corners lying on straight worldlines and edges identified
pairwise.  This reproduces the quotient space picture discussed by Mess
in the context of geometric structures \cite{Mess}.  With a different
gauge choice, Waelbroeck's model is classically
equivalent to 't Hooft's \cite{Waele}, but the two models are related
by a nonlocal change of variables, and may not be equivalent quantum
mechanically.

Much of the recent work on lattice formulations of (2+1)-dimensional
gravity have centered on spin foams and on random triangulations, both 
inherently quantum mechanical.  These will be discussed below in section
\ref{PR}.  It is worth noting here, though, that recent work on 
diffeomorphisms in spin foam models \cite{Freidel} may permit a 
classical description quite similar to that of Waelbroeck.

\subsection{Large diffeomorphisms \label{mcg}}

Up to now, I have avoided discussing an important discrete symmetry 
of general relativity on topologically nontrivial spacetimes.  The 
description of a solution of the field equations in terms of holonomies 
(sections \ref{geomstruc}--\ref{first}) or moduli (section \ref{adm})
is invariant under infinitesimal diffeomorphisms, and hence under 
``small'' diffeomorphisms, those that can be smoothly deformed to 
the identity.  But if the spacetime manifold is topologically nontrivial, 
its group of diffeomorphisms may not be connected: $M$ may admit 
``large'' diffeomorphisms, which cannot be built up smoothly from 
infinitesimal deformations.   The group of such large diffeomorphisms  
(modulo small diffeomorphisms), ${\mathcal D}(M)$, is called the 
mapping class group of $M$; for the torus $T^2$, it is also known as 
the modular group.

The archetype of a large diffeomorphism is a Dehn twist of a torus,
which may be described as the operation of cutting $T^2$ along a
circumference to obtain a cylinder, twisting one end 
by $2\pi$, and regluing.  Similar transformations exist for any
closed surface $\Sigma$, and in fact the Dehn twists around generators of 
$\pi_1(\Sigma)$ generate ${\mathcal D}(\Sigma)$ \cite{Birman,Birman2}.
It is easy to see that the mapping class group of a spacetime $M$ acts 
on $\pi_1(M)$, and therefore on the holonomies of section 
\ref{geomstruc}.  As diffeomorphisms, elements of the mapping class 
group also acts on the constant curvature metrics $\bar g_{ij}$, and 
hence on the moduli of section \ref{adm}.  

Classically, geometries that differ by actions of ${\mathcal D}(M)$ are 
exactly equivalent, so the ``true'' space of vacuum solutions for a spacetime 
with the topology $\mathbb{R}\times\Sigma$ is really $\mathcal{M}/
{\mathcal D}(M)$, where $\mathcal M$ is the moduli space (\ref{ab4}).
Quantum mechanically, it is not clear whether one should impose
mapping class group invariance on states or whether one should merely
treat ${\mathcal D}(M)$ as a symmetry under which states may transform
nontrivially (see, for instance, \cite{Isham}).  In 2+1 dimensions, though, 
there seems to be a strong argument in favor of treating the mapping 
class group as a genuine invariance, as follows.  Using the Chern-Simons 
formalism, one can compute the quantum amplitude for the scattering of 
a point particle off another particle \cite{Carlip0}, a black hole \cite{Troost},
or a handle \cite{Carlip6}.  In each case, it is only when one imposes 
invariance under the mapping class group that one recovers the correct
classical limit.  It may still be that simple enough representations of  
${\mathcal D}(M)$ lead to sensible physical results, but it is at least
clear that the mapping class group cannot be ignored.

\subsection{The torus universe \label{tor}}

The simplest nontrivial vacuum cosmology occurs for a spacetime with 
the topology $\mathbb{R}\times T^2$, where $T^2$ is 
the two-dimensional torus.  This case is in some ways exceptional---for
example, the standard metric ${\bar g}_{ij}$ of (\ref{ad4}) is flat rather
than hyperbolic---but it is also simple enough that a great deal can be
done explicitly.  Later in this review, the torus universe will 
be a canonical test of quantization; here we review classical aspects.  
The problem of finding the classical solutions, as well as an approach to 
the quantization, was, I believe, first discussed by Martinec \cite{Martinec}.  
I refer the reader to Refs.\ \cite{CarKorea,Carlipbook,CarNel,CarNel2} 
for further details.  A similarly detailed analysis may be possible
when the spatial topology is that of a Klein bottle---see, for instance,
\cite{Loukox}---but so far, this and other nonorientable examples have been
studied in much less detail.

For simplicity, let us initially restrict our attention to the case  
$\Lambda=-1/\ell^2<0$.  The group $G$ of section \ref{geomstruc}, or, 
equivalently, the gauge group in the Chern-Simons
formalism of section \ref{first}, is then $\hbox{SO}(2,2)$.  The fundamental 
group $\pi_1(\mathbb{R}\times T^2)$ has two generators, $[\gamma_1]$ 
and $[\gamma_2]$, satisfying a single relation similar to (\ref{ab5}):
\beq
[\gamma_1]\cdot[\gamma_2] = [\gamma_2]\cdot[\gamma_1] .
\label{ag1}
\eeq
The holonomy group (\ref{ab4}) is therefore generated by two commuting 
$\hbox{SO}(2,2)$ matrices, unique up to overall conjugation.  

It is a bit more convenient to describe the holonomies as elements of 
the covering group $\mathrm{SL}(2,\mathbb{R})\times\mathrm{SL}%
(2,\mathbb{R})$ \cite{NelRegZer}.  Let $\rho^\pm[\gamma_a]$ denote 
the two $\mathrm{SL}(2,\mathbb{R})$ holonomies corresponding to 
the curve $\gamma_a$.  An $\mathrm{SL}(2,\mathbb{R})$ matrix
$S$ is called hyperbolic, elliptic, or parabolic according to whether 
$|\mathop{Tr }S|$ is greater than, equal to, or less than 2, and the space 
of holonomies correspondingly splits into nine sectors.  It may be shown 
that only the hyperbolic-hyperbolic sector corresponds to a spacetime in 
which the $T^2$ slices are spacelike \cite{Ezawa,Ezawa2,LouMar,NelPic}.  
By suitable overall conjugation, the two generators of the holonomy group 
in this sector can then be taken to be
\beq
\rho^\pm[\gamma_1] = \left( \begin{array}{cc}
  e^{r_1^\pm/2} & 0 \\ 0 & e^{-r_1^\pm/2} \end{array}\right), \qquad
\rho^\pm[\gamma_2] = \left( \begin{array}{cc}
  e^{r_2^\pm/2} & 0 \\ 0 & e^{-r_2^\pm/2} \end{array}\right) ,
\label{ag2}
\eeq
where the $r_a^\pm$ are four arbitrary parameters.  Note that
this gives the right counting: the Riemann moduli space of the torus
is two dimensional, so from section \ref{adm} we expect a four-%
dimensional space of solutions.

To obtain the corresponding geometry, we can use the quotient space
construction of section \ref{geomstruc}.  Note 
first that three-dimensional anti-de Sitter space can be represented 
as the submanifold of flat $\mathbb{R}^{2,2}$ (with coordinates 
$(X_1,X_2,T_1,T_2)$ and metric $dS^2 = dX_1^2 + dX_2^2 - dT_1^2 
- dT_2^2$) defined by the condition that 
\beq
\mathop{det}|{\mathbf X}| = 1 ,
  \qquad {\mathbf X} = \frac{1}{\ell}\left( \begin{array}{cc}
  X_1+T_1 & X_2+T_2\\ -X_2+T_2 & X_1-T_1 \end{array} \right) .
\label{ag3}
\eeq
This gives an isometry between $\mathrm{AdS}_3$ and the group 
manifold of $\mathrm{SL}(2,\mathbb{R})$.  The quotient of 
$\mathrm{AdS}_3$ by the holonomy group (\ref{ag2}) may now
be obtained by allowing the $\rho^+[\gamma_a]$ to act on $\mathbf X$ 
by left multiplication and the $\rho^-[\gamma_a]$ to act by right 
multiplication.  

It is straightforward to show that the resulting induced metric is
\begin{align}
ds^2 = dt^2 &- \frac{\ell^2}{4}
  \left[(r_1^+)^2 + (r_1^-)^2 +2r_1^+r_1^-\cos\frac{2t}{\ell}\right] dx^2
  \nonumber\\  
  &- \frac{\ell^2}{2}\left[r_1^+r_2^+ + r_1^-r_2^-
  + (r_1^+r_2^- + r_1^-r_2^+)\cos\frac{2t}{\ell}\right] dxdy \label{ag4}\\
  &- \frac{\ell^2}{4}
  \left[(r_2^+)^2 + (r_2^-)^2 +2r_2^+r_2^-\cos\frac{2t}{\ell}\right] dy^2 ,
  \nonumber
\end{align}
where $x$ and $y$ are coordinates with period $1$.  An easy calculation
confirms that this is a space of constant negative curvature.  The
triad may be read off directly from (\ref{ag4}), and it is easy to solve 
(\ref{ac3}) for the spin connection $\omega$.  The resulting Chern-Simons 
connections $A^{(\pm)}$ of (\ref{ac11}) are flat, and their holonomies 
reproduce the holonomies (\ref{ag2}) of the geometric structure we began with.

To relate these expressions to the ADM formalism of section (\ref{adm}),
we must first find the slices of constant extrinsic curvature $T$.  For
the metric ({\ref{ag4}), the extrinsic curvature of a slice of constant $t$ is
$T = -\frac{2}{\ell}\cot\frac{2t}{\ell}$, which is independent of $x$ and 
$y$.  A constant $t$ slice is thus also a slice of constant York time.  The 
standard flat metric on $T^2$, the genus one version of the standard 
metric (\ref{ad4}), is
\beq
d\sigma^2 = \tau_2{}^{-1}\left| dx + \tau dy\right|^2
\label{ag5}
\eeq
where $\tau=\tau_1+i\tau_2$ is the modulus.  Comparing ({\ref{ag4}),
we see that a slice of constant $t$ has a modulus
\beq
\tau = \left(r_1^-e^{it/\ell} + r_1^+e^{-{it/\ell}}\right)
 \left(r_2^-e^{it/\ell} + r_2^+e^{-{it/\ell}}\right)^{-1} .
\label{ag6}
\eeq
The conjugate momentum $p = p^1 + ip^2$ can be similarly computed 
from (\ref{ad6}),
\beq
p= -\frac{i\ell}{ 2\sin\frac{2t}{\ell}}\left(r_2^+e^{it/\ell}
 + r_2^-e^{-{it/\ell}}\right)^2 ,
\label{ag7}
\eeq
while the ADM Hamiltonian $H$ of (\ref{ad5a}) becomes
\beq
H = \frac{\ell^2}{4}\sin\frac{2t}{\ell}(r_1^-r_2^+ - r_1^+r_2^-) 
     = \left( T^2 + \frac{4}{\ell^2}\right)^{-1/2}
         \left[\tau_2{}^2p{\bar p}\right]^{1/2} .
\label{ag8}
\eeq
In the limit of vanishing $\Lambda$, these relations go over
to those of \cite{Carlip7}.

To quantize this system, we will need the classical Poisson
brackets, which can be obtained from (\ref{ac14}): 
\beq
\{r_1^\pm,r_2^\pm\}=\mp \frac{1}{\ell}, \qquad \{r^+_a,r^-_b\}=0 .
\label{ag9}
\eeq
These, in turn, determine the brackets among the moduli and momenta 
$\tau$ and $p$,
\beq
\{\tau,\bar p\} = \{\bar\tau, p\} = 2 ,\quad
\{\tau,p\} = \{\bar\tau,\bar p\} = 0 ,
\label{ag10}
\eeq
a result consistent with (\ref{ad7}).  It may be shown that the version of
Hamilton's equations of motion coming from these brackets reproduces 
the time dependence (\ref{ag6}) of the moduli; see \cite{CarNel,Soda} for
details.  The Poisson brackets among the traces of the holonomies (\ref{ag2})
are also easy to compute.  If we let
\begin{align}
R_1^\pm &= \frac{1}{2}\mathrm{Tr}\rho^\pm[\gamma_1] 
  = \cosh\frac{r_1^\pm}{2} , \quad
R_2^\pm = \frac{1}{2}\mathrm{Tr}\rho^\pm[\gamma_2] 
  = \cosh\frac{r_2^\pm}{2} , \nonumber\\
R_{12}^\pm &= \frac{1}{2}\mathrm{Tr}\rho^\pm[\gamma_1\cdot\gamma_2]
  = \cosh\frac{(r_1^\pm+r_2^\pm)}{2} ,
\label{ag11}
\end{align}
it is not hard to check that
\beq
\{R_1^{\pm},R_2^{\pm}\}=\mp\frac{1}{{4\ell}}(R_{12}^{\pm}-
 R_1^{\pm}R_2^{\pm}) \quad \hbox{\it and cyclical permutations},
\label{ag12}
\eeq
reproducing the Poisson algebra of Nelson, Regge, and Zertuche 
\cite{NelRegZer}.

Finally, let us consider the action of the torus mapping class group.
This group is generated by two Dehn twists, which act on $\pi_1(T^2)$
by
\begin{align}
&S:\gamma_1\rightarrow \gamma_2^{-1},\hphantom{\gamma_2}
    \qquad \gamma_2\rightarrow\gamma_1\nonumber\\
&T:\gamma_1\rightarrow\gamma_1\cdot\gamma_2 ,
    \qquad \gamma_2\rightarrow\gamma_2 .
\label{ag13}
\end{align}
These transformations act on the parameters $r_a^\pm$ 
and the ADM moduli and momenta as 
\begin{xalignat}{4}
&S:r_1^{\pm}\rightarrow r_2^{\pm} & \ 
    & r_2^{\pm}\rightarrow - r_1^{\pm} & \ 
    & \tau\rightarrow -\frac{1}{\tau} & \ 
    & p\rightarrow \bar \tau^2 p \nonumber\\
&T:r_1^{\pm}\rightarrow r_1^{\pm} + r_2^{\pm} & \ 
    & r_2^{\pm}\rightarrow r_2^{\pm} & \ 
    & \tau\rightarrow \tau+1 & \ 
    &p\rightarrow p .
\label{ag14}
\end{xalignat}
These transformations are consistent with the 
relationships between the ADM and holonomy variables, and that they
preserve all Poisson brackets.

For a torus universe with zero or positive cosmological constant, similar 
constructions are possible.  I refer the reader to \cite{Carlipbook} for
details.

\subsection{Dynamics \label{dyn}}

For the torus universe of the preceding section, the dynamics can be read
off from the metric (\ref{ag4}).  The area of a slice of constant $t$ is
essentially the Hamiltonian (\ref{ag8}); it increases from $0$ at $t=0$
to a maximum at $t=\pi\ell/4$, and then shrinks to zero at $t=\pi\ell/2$.
At the ``big bang'' and ``big crunch'' the modulus (\ref{ag6}) is purely 
real, $\tau_2=0$.  This means that even apart from the ``crunch'' in volume,
the geometry is singular: a real value of $\tau$ represents a torus that has 
collapsed to a line.  For $\Lambda\ge0$, the final 
big crunch disappears, and the torus universe expands forever from an
initial big bang.  The initial spatial geometry is again degenerate.

It is not hard to check that as time increases, the modulus (\ref{ag6}) moves 
along a semicircle in the upper half of the complex plane, with a center
on the real axis.  Such a curve is a geodesic in the natural Weil-Petersson
(or Poincar{\'e}) metric on the torus moduli space \cite{HosNak,Soda}.  Because 
of the invariance under the mapping class group (\ref{ag14}), however, the 
true physical motion in the moduli space of the torus---the space of
physical configurations with the large diffeomorphisms modded out---is
much more complicated; there are arbitrarily long geodesics, and the
flow is, in fact, ergodic \cite{Cornfeld}.

For spacetimes $\mathbb{R}\times\Sigma$ with $\Sigma$ a surface
of genus $g>1$, no explicit metrics analogous to (\ref{ag4}) are known,
except for the special case of solutions with constant moduli.  The
problem is in part that no simple form such as (\ref{ag5}) for the
``standard'' constant curvature metrics exists, and in part that the
ADM Hamiltonian becomes a complicated, nonlocal function of the
moduli.  For the case of an asymptotically flat genus $g$ space, some
interesting progress has been made by Krasnov \cite{Krasnov}; I do
not know whether these methods can be extended to the spatially
closed case.  

One can write down the holonomies of the geometric structure for a
higher genus surface, of course---though even there, it is nontrivial to 
ensure that they represent spacetimes with \emph{spacelike} genus $g$ 
slices---but to a physicist, these holonomies in themselves give fairly 
little insight into the dynamics.  In principle, the ADM and Chern-Simons 
approaches might be viewed as complimentary: as Moncrief has pointed out, 
one could evaluate the holonomies in terms of ADM variables in a nice 
time-slicing, set these equal to constants, and thereby solve the ADM 
equations of motion \cite{Mon2}.  In practice, though, this approach 
seems intractable except for the genus one case.  For $\Lambda=0$, it 
may be possible to extract a useful physical picture from the geometrical 
results of Ref.\ \cite{Benedetti}, which relate holonomies to the structure 
of the initial singularity and the asymptotic future geometry, but the 
implications have not yet been explored in any depth.

A number of qualitative statements nevertheless remain possible.
The singular behavior of the torus universe carries over to 
higher genus: spacetimes with $\Lambda<0$ expand 
from a big bang and recollapse in a big crunch, while those with 
$\Lambda\ge0$ expand forever \cite{Mess,Andersson}.  Moreover,
the degeneration of the spatial geometry at the initial singularity
carries over to the higher genus case \cite{Mess,Benedetti}.  By
introducing a global ``cosmological time'' and  
exploiting recent results in two- and three-dimensional geometry,
Benedetti and Guadagnini have shown that when $\Lambda=0$,
a set of parameters describing the initial singularity and a second
set describing the geometry in the asymptotic future together
completely determine the spacetime \cite{Benedetti}.  It seems
likely that these two sets are canonically conjugate, and a better
understanding of the symplectic structure could be useful
for quantum gravity.

\section{Quantum Gravity in 2+1 Dimensions}

The reader may well have decided that for an author reviewing
quantum gravity, I have spent an inordinate amount of time
on the classical theory.  There is a good reason for this, though: each 
of the approaches described in the preceding sections leads very
naturally to an approach to quantization, which is now---with a
few twists---fairly straightforward.  Indeed, the main reason that
2+1 dimensions offer such an attractive setting for
quantum gravity is that the classical solutions can be completely
described by a finite set of parameters.  Such a description effectively 
reduces quantum gravity to quantum mechanics, allowing us to 
evade the complications of quantum field theory.  This is not 
to imply that all approaches to quantum gravity simplify---the
Wheeler-DeWitt equation, for example, apparently does not---but
it allows us to explore at least a few approaches in depth.

\subsection{Reduced phase space quantization \label{reduced}}

Perhaps the simplest approach to quantum gravity in 2+1 dimensions
\cite{Carlip7,HosNak2} begins with the reduced phase space action
(\ref{ad5}), which describes a finite-dimensional system of physical
degrees of freedom, albeit one with a complicated, time-dependent
Hamiltonian.  We know, at least in principle, how to quantize such a
system: we simply replace the Poisson brackets (\ref{ad7}) with
commutators,
\beq
[ \hat m_\alpha, \hat p^\beta ] = i\hbar\delta^\beta_\alpha ,
\label{ba1}
\eeq
represent the momenta as derivatives, $p^\alpha = -i\hbar \partial/
\partial m_\alpha$, and choose our wave functions to be square 
integrable functions $\psi(m_\alpha,T)$ that evolve according to the 
Schr\"odinger equation
\beq
i\hbar\frac{\partial\psi(m_\alpha,T)}{\partial T} = \hat H\psi(m_\alpha,T) ,
\label{ba2}
\eeq
where the Hamiltonian $\hat H$ is obtained from (\ref{ad5a}) in a
suitable operator ordering.  Invariance under the mapping class group of 
section \ref{mcg} can be incorporated by demanding that $\psi(m_\alpha,T)$ 
transform under a representation of ${\cal D}(M)$.  A similar requirement 
may help determine the operator ordering in the Hamiltonian operator 
\cite{Carlip8,Carlip9}, although some ambiguities will remain.

For spatial surfaces of genus $g>1$, the complexity of the constraint
(\ref{ad4a}) seem to make this approach to quantization impractical
\cite{Mon2}.  A perturbative expression for $\hat H$ may still exist, 
though \cite{Okamura,Okamurab}, and the Gauss map has been proposed as 
a useful tool \cite{Puzio}.  

For genus one, on the other hand, a full quantization is possible.
The classical Hamiltonian (\ref{ag8}) becomes, up to operator ordering 
ambiguities,
\beq
{\hat H} =  \left(T^2+\frac{4}{\ell^2}\right)^{-1/2}\Delta_0^{1/2} ,
\quad
\Delta_0 = -\tau_2{}^2\left( \frac{\partial^2\ }{\partial \tau_1{}^ 2} +
   \frac{\partial^2\ }{\partial \tau_2{}^ 2} \right) ,
\label{ba4}
\eeq
where $\Delta_0$ is the ordinary scalar Laplacian for the constant negative 
curvature Poincar\'e metric on moduli space, and one chooses the positive
square root in order to have a Hamiltonian that is bounded below.  This 
Laplacian is invariant under the modular transformations (\ref{ag14}), and 
its invariant eigenfunctions, the weight zero Maass forms, have been studied 
extensively by mathematicians \cite{Maass}.  The behavior of the corresponding 
wave functions has been explored by Puzio \cite{Puzio2}, who argues that 
they are well-behaved and nonsingular at the boundaries of moduli space.
Such behavior is relevant to the question of how quantum gravity handles
singularities: the degeneration of the torus geometry at the big bang,
described in section \ref{dyn}, corresponds to an approach to the boundary
of moduli space, and Puzio's results suggest that the classical singularity
may be better-behaved in the quantum theory.

A related form of quantization comes from reexpressing the moduli space for
the torus as a quotient space
$\mathrm{SL}(2,\mathbb{Z})\backslash\mathrm{SL}(2,\mathbb{R})/\mathrm{SO}(2)$
\cite{Martinec,Waldron}.  Here, the symmetric space 
$\mathrm{SL}(2,\mathbb{R})/\mathrm{SO}(2)$ describes the transverse traceless
deformations of the spatial metric, while $\mathrm{SL}(2,\mathbb{Z})$ is
the modular group.  As Waldron has observed \cite{Waldron}, this makes it
possible to reinterpret the quantum mechanical problem as that of a fictitious
free particle, with mass proportional to $\sqrt\Lambda$, moving in a quotient 
space of the (flat) three-dimensional Milne Universe.  With a suitable choice
of coordinates, though, the problem again reduces to that of understanding 
the Hamiltonian (\ref{ba4}) and the corresponding Maass forms.

While the choice (\ref{ba4}) of operator ordering is not unique, the number
of alternatives is smaller than one might expect.  The key restriction 
is diffeomorphism invariance: the eigenfunctions of the Hamiltonian should 
transform under a one-dimensional unitary representation of the mapping 
class group (\ref{ag14}).  The representation theory of this group is well-%
understood \cite{Maass2,Maass3}; one finds that the possible  
Hamiltonians are all of the form (\ref{ba4}), but with $\Delta_0$ replaced by 
\beq
\Delta_n = -\tau_2{}^2\left( \frac{\partial^2\ }{\partial \tau_1{}^2} +
  \frac{\partial^2\ }{\partial \tau_2{}^2}\right)
  + 2in\tau_2\frac{\partial\ }{\partial \tau_1} + n(n+1) , 
  \quad 2n\in\mathbb{Z} ,
\label{ba5}
\eeq
the Maass Laplacian acting on automorphic forms of weight $n$.  It has
been suggested in \cite{Carlip8} that the choice $n=1/2$ is most natural
from the point of view of Chern-Simons quantization.  Note that when 
written in terms of the momentum $p$, the operators $\Delta_n$ differ from 
each other by terms of order $\hbar$, as one would expect for operator ordering 
ambiguities.  Nevertheless, the choice of ordering may have dramatic effects 
on the physics, since the spectra of the various Maass Laplacians are quite 
different.

This ordering ambiguity may be viewed as arising from the structure of
the classical phase space.  The torus moduli space is not a manifold, but
rather has orbifold singularities, and quantization on an orbifold
is generally not unique.  Since the space of solutions of the Einstein
equations in 3+1 dimensions has a similar orbifold structure \cite{Isen},
we might expect a similar ambiguity in realistic (3+1)-dimensional
quantum gravity.

The quantization described here is an example of what Kucha{\v r} has
called an ``internal Schr\"odinger interpretation'' \cite{Kuchar}.  It 
appears to be self-consistent, and like ordinary quantum mechanics, it 
is guaranteed to have the correct classical limit on the reduced phase 
space of section \ref{adm}.  The principal drawback is that the method 
relies on a {\em classical} choice of time coordinate, which occurs as
part of the gauge-fixing needed to solve the constraints.  In particular, 
the analysis of section \ref{adm} required that we choose the York 
time-slicing from the start; a different choice might lead to a different 
quantum  theory, as it is known to do in quantum field theory \cite{Torre}.  
In other words, it is not clear that this approach to quantum gravity 
preserves general covariance.

The problem may be rephrased as a statement about the kinds of 
questions we can ask in this quantum theory.  The model naturally 
allows us to compute the transition amplitude between the spatial 
geometry of a time slice of constant mean curvature $-\mathrm{Tr}K 
= T_1$ and the geometry of a later slice of constant mean curvature  
$-\mathrm{Tr}K = T_2$.  Indeed, such amplitudes are given explicitly 
in Ref.\ \cite{Ezawa3}, where it is shown that they are peaked around
the classical trajectory.  But it is far less clear how to ask for transition 
amplitudes between other spatial slices, on which $\mathrm{Tr}K$ 
is not constant.  Such questions would seem to require a different 
classical time-slicing, and thus a different---and perhaps inequivalent%
---quantum theory.

We will eventually find a possible way out of this difficulty in section
\ref{obs}.  As a first step, we next turn to an alternative approach 
to quantization, one that starts from the first order formalism.   

\subsection{Chern-Simons quantization \label{csq}}

As we saw in section \ref{first}, (2+1)-dimensional general relativity in
first order form can be rewritten as a Chern-Simons theory.  
For compact gauge groups, the quantization of Chern-Simons theory
is well understood \cite{Wittenc,Schwarz,Schwarzb,EMSS,Axel}.  For noncompact 
groups such as those that appear in gravity, much less is known, though 
there has been some promising work \cite{Wittend,Hayashi,BarNatan,Gukov}.
Nevertheless, interesting progress can be made, especially in the simple
case of a manifold with topology $\mathbb{R}\times T^2$.

In contrast to the reduced phase space quantization of the preceding
section, our understanding of the quantum Chern-Simons gravity 
depends strongly on the sign of the cosmological constant.  For 
$\Lambda<0$, the relevant gauge group is $\mathrm{SO}(2,2)$ or its 
cover $\mathrm{SL}(2,\mathbb{R})\times\mathrm{SL}(2,\mathbb{R})$.
This is the most poorly understood case; an explicit quantization of the
algebra holonomies exists for genus one (see below) and genus two
\cite{NelReg2}, but more general results do not yet exist.  

For $\Lambda>0$, the relevant gauge group is $\mathrm{SO}(3,1)$ or 
its cover $\mathrm{SL}(2,\mathbb{C})$, a complex gauge group whose
Chern-Simons theory is somewhat better understood \cite{Wittend,%
Hayashi,BarNatan}.  As noted in section \ref{first}, the Poisson brackets
for this theory are related to the quantum double of the Lorentz group,
and Buffenoir et al.\ have used this structure to write down an explicit 
quantization \cite{Buff}.  As far as I know, the relationship between 
this work, which is based on a Hamiltonian formalism and combinatorial 
quantization, and that of Witten and Hayashi \cite{Wittend,Hayashi}, 
which is based on geometric quantization, has not yet been explored.

For $\Lambda=0$, the relevant gauge group is $\mathrm{ISO}(2,1)$,
the (2+1)-dimensional Poincar{\'e} group, or its universal cover.  Here
there is again a connection to the quantum double of the Lorentz
group, which has been used in \cite{Bais,Bais2,Meusburger} to explore 
the quantum theory, although largely in the context of point particles.
In this case, one has the nice feature that the phase space has a natural
cotangent bundle structure, allowing us to immediately identify the 
holonomies of the spin connection $\omega$ as generalized positions,
and their derivatives as generalized momenta.  This provides a direct
link to the loop variables of Ashtekar, Rovelli, and Smolin \cite{Ashtekar,%
Ashtekarb},
\beq
T^0[\gamma] = \frac{1}{2}\mathrm{Tr}\rho_0[\gamma,x], \quad
T^1[\gamma] =  \int_\gamma 
  \mathrm{Tr}\left\{ \rho_0[\gamma, x(s)]\, e^a(\gamma(s)){\mathcal J}_a \right\} ,
\label{bb1}
\eeq
where
\beq
\rho_0[\gamma,x] = P\exp\left\{\int_\gamma \omega^a{\mathcal J}_a \right\}
\label{bb2}
\eeq
is the $\mathrm{SL}(2,\mathbb{R})$ holonomy of the spin connection and  
$T^1[\gamma]$ can be expressed as a derivative of $T^0]\gamma]$ along a 
path in the space of flat connections \cite{Carlipbook}.  Note that the 
generator $\mathcal{J}$ may, in principle, be in any representation of 
$\mathrm{SL}(2,\mathbb{R})$, and that the trace in (\ref{bb1}) may depend 
on the choice of representation.  I will return to the resulting quantum 
theory, loop quantization, in section \ref{loopvar}.

As in reduced phase space quantization, matter simplify considerably for 
the torus universe $\mathbb{R}\times T^2$.  Let us again focus
on the case $\Lambda<0$.  A complete---in fact, overcomplete---set of 
observables is given by the traces (\ref{ag11}) of the holonomies, and 
our goal is to quantize the algebra (\ref{ag12}).  To do so, we proceed 
as follows:
\begin{enumerate}
\item We replace the classical Poisson brackets $\{\,,\,\}$ by commutators
$[\,,\,]$, with the rule
$
[x,y]= xy-yx =i \hbar \{x,y\} ;
$
\item On the right hand side of (\ref{ag12}), we replace the product by
the symmetrized product,
$
xy \to \frac{1}{2} (xy +yx) .
$
\end{enumerate}
The resulting algebra is defined by the relations
\beq
\hat R_1^{\pm}\hat R_2^{\pm}e^{\pm i \theta}
  - \hat R_2^{\pm}\hat R_1^{\pm} e^{\mp i \theta}=
  \pm 2i\sin\theta\, \hat R_{12}^{\pm} \quad \hbox{\it and cyclical
  permutations}
\label{bb3}
\eeq
with $\tan\theta= -{\hbar/8\ell}$.  The algebra (\ref{bb3}) is not a 
Lie algebra, but it is related to the Lie algebra of the quantum group 
$U_q(\mathit{sl}(2))$ with $q=\exp{4i\theta}$ \cite{NelRegZer}.  Classically,
the observables $R_1^\pm$, $R_2^\pm$, and $R_{12}^\pm$ are not independent;
in the quantum theory, the corresponding statement is that the quantities
\beq
\hat F^{\pm}
  = 1-{\tan}^2\theta - e^{\pm 2i\theta} \left( (\hat R_1^\pm)^2 
  + (\hat R_{12}^\pm)^2\right) 
  - e^{\mp 2i\theta} (\hat R_2^\pm)^2
  + 2e^{\pm i\theta}\cos\theta \hat R_1^\pm \hat R_2^\pm \hat R_{12}^\pm 
\label{bb3a}
\eeq
commute with the holonomies, and can be consistently set to zero.
In terms of the parameters $r_a^\pm$ of (\ref{ag2}), the algebra 
can be represented by \cite{CarNel,CarNel2}
\beq
\hat R_1^{\pm} = \sec\theta\cosh\frac{\hat r_1^{\pm}}{2} ,\quad
\hat R_2^{\pm} = \sec\theta\cosh\frac{\hat r_2^{\pm}}{2} ,\quad
\hat R_{12}^{\pm} = \sec\theta\cosh\frac{(\hat r_1^{\pm}+\hat r_2^{\pm})}{ 2}
\label{bb4}
\eeq
with
\beq
[\hat r_1^{\pm}, \hat r_2^{\pm}] = \pm 8i\theta,
  \qquad[\hat r^+_a,\hat r^-_b] = 0 .
\label{bb5}
\eeq
For $\Lambda$ small, these commutators differ from the naive quantization
of the classical brackets (\ref{ag9}) by terms of order $\hbar^3$.  An
alternative quantization, also differing by terms of order $\hbar^3$,
works directly with the holonomy matrices (\ref{ag2}), imposing a 
quantum-group-like quantization condition \cite{NelPicb}
\beq
\rho^+[\gamma_1]\rho^+[\gamma_2] = q \rho^+[\gamma_2]\rho^+[\gamma_1] ,
\quad q = e^{-i\hbar/4\ell}
\label{bb5a}
\eeq
with a similar condition for $\rho^-$.

We must also implement the action of the modular group (\ref{ag14})
on the operators $\hat R^\pm_a$.  One can find an action preserving the
algebraic relations (\ref{bb3}), corresponding to a particular factor 
ordering of the classical modular group.  The Nelson-Picken 
quantization (\ref{bb5a}) admits a similar modular group action.

For a full quantum theory, of course, one needs not only an abstract
operator algebra, but a Hilbert space upon which the operators act.
For the $\mathbb{R}\times T^2$ universe, (\ref{bb5}) suggests that
a natural choice is to take wave functions to be square integrable
functions of the $r_2^\pm$.  There is a potential difficulty here,
however: the modular group does not act properly discontinuously
on this configuration space.  This means that the quotient of this
space by the modular group is badly behaved;  in fact, there are
no nonconstant modular invariant functions of the $r_2^\pm$
\cite{LouMar,Giulini,Peldan}.  We shall return to this problem in
the section \ref{obs}.
 
\subsection{Covariant canonical quantization \label{ccq}}

The technique of Chern-Simons quantization relies on special features of 
general relativity in 2+1 dimensions, and does not readily generalize to 
higher dimensions.  It is, however, closely related to a much more 
general approach, covariant canonical quantization \cite{Ashtekarc,%
Ashtekard,Crnkovic,Lee,Wald}, or ``quantization of the space of classical 
solutions.'' 

Our starting point is the observation that the phase space of a well-behaved 
classical theory is isomorphic to the space of classical solutions.  Indeed,
if $\mathcal C$ is an arbitrary Cauchy surface, then a point in the phase 
space determines initial data on $\mathcal C$, which can be evolved to
give a unique solution, while, conversely, a classical solution restricted to 
$\mathcal C$ determines a point in the phase space.  Moreover, the space 
of solutions has natural symplectic structure \cite{Lee,Wald}, which can 
be shown to be equivalent to the standard symplectic structure on phase
space.  For the case of (2+1)-dimensional gravity, this equivalence is
demonstrated in section 6.1 of \cite{Carlipbook}.

For (2+1)-dimensional gravity, the space of classical solutions is the 
space of geometric strictures of section \ref{geomstruc}.  If we restrict 
our attention to spacetimes with the topology $\mathbb{R}\times\Sigma$
with $\Sigma$ closed and $\Lambda\le0$, the holonomies of a geometric 
structure determine a unique maximal domain of dependence \cite{Mess}, 
exactly the right setting for covariant canonical quantization.  But 
as we saw in section \ref{first}, the holonomies of a geometric structure 
are precisely the holonomies of the Chern-Simons formalism, and the 
symplectic structures are the same as well.  Thus in this setting, 
Chern-Simons quantization \emph{is} covariant canonical quantization.  
If $\Lambda>0$ or point particles are present, the holonomies do not quite
determine a unique geometric structure, and the Chern-Simons theory 
is not quite equivalent to general relativity.  In that case, additional 
discrete variables might be necessary; see, for example, \cite{Ezawa} 
for the case of a torus universe with $\Lambda>0$.

As we shall see in section \ref{obs}, the construction of dynamical 
observables and time-dependent states in covariant canonical quantum theory 
requires an explicit isomorphism between the phase space and the space of 
classical solutions.  For the torus universe, such an isomorphism is known.  
For higher genus spaces, however---and certainly for realistic (3+1)-dimensional 
gravity---it is not \cite{Mon2}.  Often, however, we can determine such an 
isomorphism perturbatively in the neighborhood of a known classical solution.  
This raises the interesting question, so far answered only in simple models
\cite{Basu}, of whether classical perturbation theory can be used to define 
a perturbative covariant canonical quantum theory.

\subsection{A digression: observables and the problem of time \label{obs}}

When one attempts to interpret the quantum theories coming from the
Chern-Simons formalism or covariant canonical quantization, one finds an 
immediate and rather profound difficulty.  The gauge-invariant observables---the 
traces of the holonomies---are automatically nonlocal and time-independent, 
and one obtains a ``frozen time formalism,'' or what Kucha{\v r} has called 
``quantum gravity without time'' \cite{Kuchar}.  In one sense, this is a 
good thing: one knows from general arguments that the diffeomorphism-invariant 
observables in any quantum theory of gravity must have these features 
\cite{Torreb}.  On the other hand, it is not at all easy to see how to 
extract local geometry and dynamics from such a picture: if our only
observables are nonlocal and time-independent, how can we recover a classical
limit with local excitations that evolve in time?

Quantum gravity in 2+1 dimensions offers a possible answer to this
dilemma.  Note first that the problem is already present classically.  A 
geometric structure determines a spacetime, and must contain within
it all of the dynamics of that spacetime.  On the other hand, the basic 
data that fix the geometric structure---the transition functions, or, often, 
the holonomies---have no obvious dynamics.  In principle, the classical
answer is simple:  
\begin{enumerate}
\item Use, say, the holonomies to determine a spacetime geometry;
\item Select a favorite time-slicing;
\item Read off the spatial metric and its time derivatives from the
  spacetime metric of step 1 in this slicing.
\end{enumerate}
This procedure can be understood as a concrete realization of the 
isomorphism described in section \ref{ccq} between the phase space 
and the space of classical solutions, with the Cauchy surface $\mathcal C$
fixed by the choice of time-slicing.

For the simple case of the torus universe, these steps can be transcribed
almost directly to the quantum theories.  Equations (\ref{ag6})--(\ref{ag8})
become definitions of operators,
\begin{align}
{\hat\tau}_t &= \left({\hat r}_1^-e^{it/\ell} + {\hat r}_1^+e^{-{it/\ell}}\right)
 \left({\hat r}_2^-e^{it/\ell} + {\hat r}_2^+e^{-{it/\ell}}\right)^{-1} \nonumber\\
{\hat p}_t &= -\frac{i\ell}{ 2\sin\frac{2t}{\ell}}\left({\hat r}_2^+e^{it/\ell}
 + {\hat r}_2^-e^{-{it/\ell}}\right)^2 \label{bd1}\\
{\hat H}_t &= \frac{\ell^2}{4}\sin\frac{2t}{\ell}
  ({\hat r}_1^-{\hat r}_2^+ - {\hat r}_1^+{\hat r}_2^-) ,\nonumber
\end{align}
where the operator ordering has been chosen to respect the modular 
transformations (\ref{ag14}).  The parameter $t$ is now merely a 
label for a one-parameter family of diffeomorphism-invariant observables.  
These observables obtain their physical significance from the classical
limit: ${\hat\tau}_t$, for example, is the operator whose expectation
value gives the mean value of the modulus on a time slice of constant mean
curvature $T = -\frac{2}{\ell}\cot\frac{2t}{\ell}$.  Such observables
are examples of what Rovelli has called ``evolving constants of motion''
\cite{Rovellib,Rovellic}.

From this point of view, we should think of Chern-Simons/covariant canonical 
quantization as a sort of Heisenberg picture, with time-independent
states and ``time''-dependent operators.  To obtain the corresponding
Schr{\"o}dinger picture, we proceed as in ordinary quantum mechanics:
we diagonalize ${\hat\tau}_t$, obtaining a transition matrix 
$K(\tau,{\bar\tau};t | r_2^+,r_2^-) 
=  \langle \tau,{\bar\tau}; t | r_2^+,r_2^-\rangle$ that allows us to
transform between representations \cite{Carlip8,CarNel3}.  The resulting
``time''-dependent wave functions obey a  Schr{\"o}dinger
equation of the form (\ref{ba2})--(\ref{ba4}), but with the Laplacian in
$\hat H$ replaced by the weight $1/2$ Maass Laplacian $\Delta_{1/2}$
of (\ref{ba5}).  In Ref.\ \cite{Ezawa3}, it has been shown that these wave 
functions are peaked around the correct classical trajectories.  (Different 
operator orderings in (\ref{bd1}) give different weight Laplacians 
\cite{Carlip9}.)

As a useful byproduct, this analysis allows us to solve the problem of 
the poorly-behaved action of the modular group discussed at the end 
of section \ref{csq} \cite{CarNel3,CarNel4}.  If we start with a reduced 
phase space wave function $\tilde\psi(\tau,{\bar\tau};t)$ and use the 
transition matrix $K$ to determine a Chern-Simons wave function 
$\psi( r_2^+,r_2^-)$, we find, indeed, that $\psi( r_2^+,r_2^-)$ is not 
modular invariant.  Instead, though, the entire Hilbert space of Chern-Simons
wave functions splits into ``fundamental regions,'' orthogonal 
subspaces that transform into each other under the action of the
modular group.  Any one of these fundamental regions is equivalent
to any other, and each is equivalent to the Hilbert space arising from
reduced phase space quantization.  Moreover, matrix elements of
any modular invariant function vanish unless they are taken between
states in the same fundamental region.  Modular invariance thus takes 
a slightly unexpected form, but can still be imposed by restricting the
theory to a single fundamental region of the Hilbert space.

We can also begin to address the problem raised at the end of section
\ref{reduced}, the limited and slicing-dependent range of questions
one can ask in reduced phase space quantization.  The operators
(\ref{bd1}) introduced here on the covariant canonical Hilbert space
were obtained from a particular classical time-slicing, and answer
questions about spatial geometry in that slicing.  In principle, however, 
we can choose any other slicing, with a new time coordinate $\bar t$,
and determine the corresponding operators ${\hat\tau}_{{\bar t}}$,
${\hat p}_{{\bar t}}$, and ${\hat H}_{{\bar t}}$.  The operator ordering
of such operators will, of course, be ambiguous, though one might
hope that the action of the modular group might again restrict the 
choices.  But such an ambiguity need not be seen as a problem with the 
theory; rather, it is merely a statement that many different quantum 
operators can have the same classical limit, and that ultimately experiment 
must decide which operator we are really observing.

There is, to be sure, a danger that the ``Schr{\"o}dinger pictures''
coming from different time-slicings may not be consistent.  Suppose,
for example, that we choose two slicings that agree on an initial and
a final slice $\Sigma_1$ and $\Sigma_2$, but disagree in between.  
If we start with an initial wave function on $\Sigma_1$, we must
check that the Hamiltonians coming from the different slicings
evolve us to the same final wave function on $\Sigma_2$.  For field
theories, even in flat spacetime, this will not always happen \cite{Torre}.
For (2+1)-dimensional gravity, on the other hand, there is evidence
that one can always find operator orderings of the Hamiltonians that 
ensure consistent evolution \cite{Cosgrove}.  If this ultimately turns
out not to be the case, however, it may simply mean that we should
treat the covariant canonical picture as fundamental, and discard
the Schr{\"o}dinger pictures of time-dependent states.

\subsection{``Quantum geometry'' \label{loopvar}}

We now resume the discussion of alternative approaches to 
quantum gravity in 2+1 dimensions.  In 3+1 dimensions, one of the
most attractive programs of quantization is ``loop quantum gravity,''
or ``quantum geometry'' \cite{Ashtekar,Rovellid}.  In 2+1 dimensions
with $\Lambda=0$, this approach is closely related to the first 
order formalism of section \ref{first}, but takes as its fundamental 
observables the loop variables $T^0[\gamma]$ and $T^1[\gamma]$ 
of (\ref{bb1}).  More precisely, loop quantum gravity starts with a 
Hamiltonian formulation of the first order formalism, with constraints,
written in analogy to the (3+1)-dimensional case \cite{Ashtekar}, that 
take the form
\beq
D_i{\tilde E}^{ia} =0, \quad {\tilde E}^i{}_aR^a_{ij} = 0 \quad
\epsilon_{abc}{\tilde E}^{ib}{\tilde E}^{jc}R^a_{ij} = 0 .
\label{be0}
\eeq
Here, indices $i,\,j,\,k$ are spatial indices on a surface of 
constant time, ${\tilde E}^{ia} = \epsilon^{ij}e_j{}^a$, $D_i$ is the 
$\mathrm{SO}(2,1)$ gauge-covariant derivative for the connection 
$\omega$, and the $R^a_{ij}$ are the spatial components of the curvature 
two-form (\ref{ac4}).  When the spatial metric $g_{ij}=e_i{}^ae_{ja}$ 
is nondegenerate, it is straightforward to show that these constraints 
are equivalent to the standard constraints of first order gravity, and 
the classical theories are identical.  When $g_{ij}$ is noninvertible, 
on the other hand, the constraints are not equivalent.  Instead, the 
constraints (\ref{be0}) yield a phase space with infinitely many 
degrees of freedom, arising from the geometries formed from an arbitrary
collection of independent patches of ordinary spacetime separated 
by regions with degenerate metrics \cite{Varadarajan,Varadarajanb}.  
Implications of such degenerate configurations for the quantum 
theory are not well-understood.

Let us restrict ourselves to invertible spatial metrics, and attempt 
to quantize the algebra of loop variables ${\widehat T}^0[\gamma]$ 
and ${\widehat T}^1[\gamma]$.  For the torus universe, 
it is not hard to show that such a quantization simply reproduces the 
theory we already obtained in the Chern-Simons formulation
(see, for example, section 7.2 of \cite{Carlipbook}).  So far, there is
nothing new here.

There is an another way to look at the operator algebra of the
operators ${\widehat T}^0[\gamma]$ and ${\widehat T}^1[\gamma]$,
however, which leads to a new approach, the loop representation. 
Up to now, we have been thinking of the operators ${\widehat T}$
as a set of functions of the triad and spin connection, indexed by 
loops $\gamma$.  Our wave functions are thus functionals of the
``configuration space'' variable $\omega$, or, more precisely, 
functions on the moduli space of flat $\mathrm{SO}(2,1)$ or
$\mathrm{SL}(2,\mathbb{R})$ connections on $\Sigma$.  But we 
could equally well view the ${\widehat T}$ operators as functions of 
loops---or in 2+1 dimensions, homotopy classes $[\gamma]$ of 
loops---indexed by $e$ and $\omega$.  Wave functions would 
then be functions of loops or sets of loops.  This change of 
viewpoint is rather like the decision in ordinary quantum 
mechanics to view a wave function $e^{ipq}$ as a function on 
momentum space, indexed by $q$, rather than a function on 
position space, indexed by $p$.   

The loop representation is complicated by the existence of Mandelstam
identities \cite{Mandelstam} among holonomies of loops, but for the
case of the torus universe, a complete, explicit description of the states 
is again possible \cite{Ashtekar,Ashtekarb}.  The simplest construction
begins with a vacuum state $|0\rangle$ annihilated by every operator
${\widehat T}^1[\gamma]$, and treats the ${\widehat T}^0[\gamma]$
as ``creation operators.''  Since any homotopy class $[\gamma]$ of
loops on the torus is completely characterized by a pair of winding
numbers $(m,n)$, one can write these states as $|m,n\rangle$.  The
action
\begin{align}
{\widehat T}^0[m,n] |p,q\rangle &= \phantom{-} \frac{1}{2} \left(
   |m+p,n+q\rangle + |m-p,n-q\rangle\right) \nonumber\\
{\widehat T}^1[m,n] |p,q\rangle &= -\frac{i}{8} (mq-np) \left(
   |m+p,n+q\rangle - |m-p,n-q\rangle\right) 
\label{be1}
\end{align}
then gives a representation of the loop algebra.

Observe now that the loop variables $T^0[\gamma]$ depend only
on the ``configuration space'' variable $\omega$.  We can thus
relate the loop representation to the Chern-Simons 
representation by simultaneously diagonalizing these operators, 
obtaining wave functions that are functions of the $\mathrm{SO}(2,1)$
holonomies alone.  For the torus universe, this ``loop transform'' can
be obtained explicitly \cite{Ashtekar,Ashtekarb,Marolf}, and written
as a simple integral transform.

The properties of this transform depend on the holonomies, that is,
the eigenvalues of ${\widehat T}^0[\gamma]$.  For simplicity, let us take 
the generator $\mathcal{J}$ in (\ref{bb2}) to be in the two-dimensional
representation of $\mathrm{SL}(2,\mathbb{R})$.  In the ``timelike
sector,'' in which the traces of the two holonomies are both less than 
two, the loop transform is a simple Fourier transformation, and 
Chern-Simons and loop quantization are equivalent.  

Unfortunately, though, this is not the physically relevant sector: it does 
not correspond to a geometric structure with spacelike $T^2$ slices.  
For a physically interesting geometry, one must go to the ``spacelike 
sector,'' in which the traces of the holonomies are both greater than 
two.  In this sector, the transform is {\em not\/} very well-behaved: in 
fact, a dense set of Chern-Simons states transforms to zero \cite{Marolf}.  
The loop representation thus appears to be rather drastically different 
from the Chern-Simons formulation.

The problems in the physical sector can be traced back to the fact that 
$\mathrm{SL}(2,\mathbb{R})$ is a noncompact group.  There have 
been two proposals for an escape from this dilemma.  One is to start 
with a different dense set of Chern-Simons states that transform faithfully, 
determine the inner product and the action of the $\widehat T$ operators 
on the resulting loop states, and then form the Cauchy completion to 
define the Hilbert space in the loop representation \cite{Marolf}.  This 
is a consistent procedure, but many of the resulting states in the Cauchy 
completion are no longer functions of loops in any clear sense; they 
correspond instead to ``extended loops'' \cite{Gambini}, whose 
geometrical interpretation is not entirely clear.  A second possibility is 
to change the integration measure in the loop transform to make various 
integrals converge better \cite{AshLoll}.  Such a choice introduces order 
$\hbar$ corrections to the action of the ${\widehat T}^1$ operators, and 
one must be careful that the algebra remains consistent.  This is possible,
but at some cost---the inner products between loop states become 
considerably more complex, as does the action of the mapping class 
group---and it is not obvious that there is a canonical choice of
the new measure and algebra.

A third possibility is suggested by recent work on spin networks for 
noncompact groups \cite{Freidel2,Freidel3}.  This new technology 
essentially allows one to consider holonomies  (\ref{bb2}) that lie in
infinite-dimensional unitary representations of the Lorentz group,
with a finite inner product defined by appropriate gauge-fixing.
The quantities $T^0$ and $T^1$ can be represented as Hermitian
operators on this space of holonomies (or on a larger space of spin
networks).  At this writing, implications of this approach for the
loop transform in 2+1 dimensions have not yet been investigated.

Finally, I should briefly mention the role of spin networks in 
(2+1)-dimensional quantum geometry.  In the (3+1)-dimensional
theory, loop states have been largely superseded by spin network
states, states characterized by a graph $\Gamma$ with edges
labeled by representations and vertices labeled by intertwiners
\cite{Rovellid}.  Such states can be defined in 2+1 dimensions as
well, and there has been some interesting recent work on their role
as ``kinematic'' states \cite{Freidel3}.  In 2+1 dimensions, however,
the full constraints imply that such states have their support on flat
connections, and only holonomies around noncontractible curves describe 
nontrivial physics.  An interesting step toward projecting out the
physical states has recently been taken in \cite{Noui}, in the
context of Euclidean quantum gravity; the ultimate effect is to reduce 
spin network states to loop states of the sort we have considered above.
A better understanding of the relationship to the gauge-fixing procedure  
of Refs.\ \cite{Freidel2,Freidel} would be valuable.   

\subsection{Lattice methods I: Ponzano-Regge and spin foams\label{PR}}

A long-standing approach to quantum gravity in 3+1 dimensions has been to
look for discrete approximations to the path integral \cite{Lollrev,WilReg}: 
quantized Regge calculus \cite{Regge0}, for example, or sums over random
triangulations \cite{Ambjorn}.  The basic idea is that although the
full ``sum over geometries'' may be impossible to evaluate, a sum over 
appropriately discretized geometries might give a good approximation,
perhaps becoming extremely good near a phase transition at which lattice
spacing can go to zero.  When applied to 2+1 dimensions, such
methods have the added feature of sometimes being exact: since
geometries satisfying the constraints have constant or zero curvature,
a discrete ``approximation'' may give a complete description.

Regge calculus in 2+1 dimensions begins with a triangulated three-manifold, 
consisting of a collection of flat simplices joined along one-dimensional 
edges.  The curvature of such a manifold is concentrated entirely at the
edges.  For a simplicial three-manifold with Riemannian signature, composed 
of simplices with edges of length $l_e$, Regge's form of the Einstein-Hilbert 
action is
\beq
I_{\mathit{Regge}}
   = 2\sum_{\hbox{\scriptsize edges:$e$}}\delta_e \ell_e ,
\label{bf1}
\eeq
where $\delta_e$ is the conical deficit angle at the edge labeled by 
the index $e$.  A similar expression exists for Lorentzian signature, 
although the definition of the deficit angle is a bit more complicated
\cite{Barrett}. 

The first hint that (2+1)-dimensional gravity might be exceptional came 
from the observation by Ponzano and Regge \cite{Ponzano} that the Regge
action in 2+1 dimensions can be re-expressed in terms of Wigner-Racah 
$6j$-symbols.  (See \cite{Carter} for more about these quantities.)  
Consider first a single tetrahedron with edge lengths $\ell_i = 
\frac{1}{2}(j_i+\frac{1}{2})$, where the $j_i$ are integers or 
half-integers.  Ponzano and Regge noticed, and Roberts later proved
rigorously \cite{Robertsa}, that in the limit of large $j$,
\begin{multline}
\lefteqn{\exp\{\pi i\sum_{i=1}^6 j_i\} \left\{ 
  \begin{array}{ccc} 
  j_1 & j_2 & j_3\\ j_4 & j_5 & j_6 \end{array} \right\}
  \sim} \\
\frac{1}{\sqrt{6\pi V}}\left[ 
  \exp\left\{i\left(I_{\mathit{Regge}}+\frac{\pi}{4}\right)
  \right\} + 
  \exp\left\{-i\left(I_{\mathit{Regge}}+\frac{\pi}{4}\right)
  \right\}  \right] ,
\label{bf2}
\end{multline}
where $\left\{ \begin{array}{ccc} j_1 & j_2 & j_3\\ j_4 & j_5 & j_6 
\end{array} \right\}$ is a $6j$ symbol, $I_{\mathit{Regge}}$ is 
the Regge action (\ref{bf1}) for the tetrahedron, and $V$ is its 
volume. For a manifold made of a collection of 
such tetrahedra, the full Regge action will occur in a product of
such $6j$ symbols.  This suggests that the (2+1)-dimensional path 
integral, which is essentially a sum over geometries of terms of the 
form $\exp\left\{-iI_{\mathit{Regge}}\right\}$, might be expressible 
as a sum of such products.  Ponzano and Regge's specific proposal, 
developed by Hasslacher and Perry \cite{Hasslacher} and modified
by Ooguri \cite{Ooguri} to account for boundaries, was the following:  

Consider a three-manifold $M$ with boundary $\partial M$, with a given
triangulation $\Delta$ of $\partial M$.  Choose a triangulation of $M$ 
that agrees with the triangulation of the boundary.  Label interior 
edges of tetrahedra by integers or half-integers $x_i$ and exterior 
(boundary) edges by $j_i$, and for a given tetrahedron $t$, let $j_i(t)$ 
denote the spins that color its (interior and exterior) edges.  
Then
\begin{multline}
Z_{\Delta}[\{j_i\}] = \lim_{L\rightarrow\infty} 
  \sum_{\hbox{\scriptsize $x_e\le L$}} \Biggl(
  \prod_{\hbox{\scriptsize ext.\,edges:$i$}}(-1)^{2j_i}\sqrt{2j_i+1}
    \prod_{\hbox{\scriptsize int.\,vertices}} \Lambda(L)^{-1} \\
  \prod_{\hbox{\scriptsize int.\,edges:$\ell$}} (2x_\ell+1)
  \prod_{\hbox{\scriptsize tetra:$t$}} (-1)^{\sum_{i=1}^6j_i(t)}
  \left\{ \begin{array}{ccc} 
  j_1(t) & j_2(t) & j_3(t)\\ j_4(t) & j_5(t) & j_6(t) \end{array} \right\}
  \Biggr) ,
\label{bf3}
\end{multline}
where ``int'' and ``ext'' mean ``interior'' and ``exterior'' and
\beq
\Lambda(L) = \sum_{j\le L} (2j+1)^2
\label{bf4}
\eeq
is a regularization factor that controls divergences in the sum over
interior lengths.  With this weighting, identities among 
$6j$-symbols may be used to show that the amplitude is 
invariant under refinement---that is, subdivision of a tetrahedron  
into four smaller tetrahedra---suggesting that we are dealing with a 
``topological'' theory that does not depend on the choice of
triangulation.  This is, of course, what one would hope for, based 
on the classical characteristics of (2+1)-dimensional gravity.  

The ``topological'' feature of the Ponzano-Regge model was made more precise 
by Turaev and Viro \cite{Turaev}, who discovered an improved regularization, 
based on the technology of quantum groups.  The ``spins'' $j$ in (\ref{bf3}) 
can be viewed as labeling representations of $\mathrm{SU}(2)$.  If these 
are replaced by representations of the quantum group $U_q(\mathit{sl}(2))$ 
(``quantum $\mathrm{SU}(2)$''), with $q = \exp\left\{\frac{2\pi i}{k+2}\right\},\
k\in\mathbb{Z}$, the number of such representations is finite, and the sum 
over interior edge lengths is automatically cut off.  With appropriate
substitutions (e.g., ``quantum'' $6j$ symbols \cite{Carter}), the
Ponzano-Regge amplitude (\ref{bf3}) becomes well-defined without any
regularization.

The construction of physical states as appropriate functions of boundary
edge lengths is described in section 11.2 of \cite{Carlipbook}.  The
resulting amplitudes can be computed for simple topologies \cite{Ion,Ionb},
and have several key features:
\begin{enumerate}
\item For large but finite $k$, the Turaev-Viro quantum group regularization  
introduces a cosmological constant to the Regge action \cite{Mizoguchi,Mizoguchib}
\beq
\Lambda = \left( \frac{4\pi}{k}\right)^2 .
\label{bf5}
\eeq
Correspondingly, the quantum $6j$ symbols are related to spherical tetrahedra
rather than flat tetrahedra \cite{Taylor}.
\item In the large $k$ limit, the Turaev-Viro Hilbert space is isomorphic 
to the space of gauge-invariant functions of flat $\mathrm{SU}(2)$ connections 
\cite{Ooguri,Oogurib,Sasakura}.  This establishes a direct link to Chern-Simons 
quantization: just as (2+1)-dimensional Lorentzian gravity can be written 
as an $\mathrm{ISO}(2,1)$ Chern-Simons theory with a configuration space of 
flat $\mathrm{SO}(2,1)$ connections, three-dimensional Euclidean gravity can 
be written as an $\mathrm{ISU}(2)$ Chern-Simons theory with a configuration 
space of flat $\mathrm{SU}(2)$ connections.
\item For a closed three-manifold $M$, the Turaev-Viro amplitude---now
interpreted as a partition function---is equal to the absolute square of the 
partition function of an $\mathrm{SU}(2)$ Chern-Simons theory with coupling 
constant $k$ \cite{Rama,Turaevc,Roberts},
\beq
Z_{\mathit{TV}} = |Z_{\mathit{CS}}|^2 .
\label{bf6}
\eeq
This again establishes an equivalence with Euclidean gravity in first-order 
form: the first-order Euclidean action with $\Lambda>0$ can be written as 
a difference of $\mathrm{SU}(2)$ Chern-Simons actions, so
\beq
Z_{\mathit{grav}} = \int [dA^+][dA^-] e^{i(I[A^+] - I[A^-])} = 
  \left| \int [dA^+]e^{iI[A^+]} \right|^2 ,
\label{bf7}
\eeq
in agreement with (\ref{bf6}).
\item A candidate for a discrete version of the Wheeler-DeWitt equation 
in three dimensions has been found \cite{Barrettb}, for which the 
Ponzano-Regge wave functions are solutions.
\end{enumerate}

Although it has not been universally appreciated, the existence of a 
divergence in the sum (\ref{bf3})---regulated either by an explicit cut-off 
or by quantum group tricks---is rather mysterious, given the absence of
local excitations and the general well-behavedness of gravity in three
dimensions.  This mystery may have recently been solved by Freidel and 
Louapre \cite{Freidel}, who show that a residual piece of the diffeomorphism 
symmetry has not been factored out of the Ponzano-Regge action.  Because of 
this symmetry, the sum (\ref{bf3}) overcounts physical configurations, and 
the regulator $\Lambda(L)$ is simply the remaining gauge volume.  Freidel 
and Louapre further show that the symmetry can instead be 
gauge-fixed, leading to a sum over a restricted and considerably simplified 
class of ``collapsed'' triangulations. 

While the mathematics of Ponzano-Regge and Turaev-Viro models has been 
studied extensively, so far only a bit of attention has been given to the
``traditional'' issues of quantum gravity.  A few numerical investigations 
of the Ponzano-Regge path integral have been undertaken \cite{Hamber}, 
but the evidence of a continuum limit is thus far inconclusive.  The model 
has been used to study conditional probabilities and the emergence of 
quasiclassical behavior in quantum gravity \cite{Petryk}, but the 
cut-off dependence of these results makes their physical significance 
unclear.  In an interesting recent paper, Colosi et al.\ have investigated
the dynamics of a single tetrahedron \cite{Colosi}, showing that a quantum
description of the evolution can be given in terms of a boundary amplitude.

A number of observables, whose expectation values generally give topological 
information about the spacetime or about knots within spacetime, have been 
discussed in \cite{Archer,Beliakova,Turaev1,Garcia1}.  With a few 
exceptions, though, work in this area has remained largely mathematical in 
nature; fairly little is understood about the physics of these observables, 
although some are probably related to length spectra \cite{Barrettc} and 
perhaps volumes \cite{FreiKras,Carbone}, and others are almost certainly 
connected to scattering amplitudes for test particles.   

The Ponzano-Regge and Turaev-Viro models are examples of ``spin foam''
models \cite{Baez,Perez}, that is, a model based on  simplicial complexes 
with faces, edges, and vertices labeled by group representations and 
intertwiners.  A key question is whether one can extend such models to 
Lorentzian signature.  It has been known for several years how to generalize 
the Ponzano-Regge action for a single tetrahedron \cite{Barrett,Davids,Garcia}, 
and recently considerable progress has been made in constructing Lorentzian 
spin foam models \cite{Perez,Freidelb,Davidsb}.  

Probably the most elegant derivation of a Lorentzian spin foam description
starts with the first-order action (\ref{ac2}), with $\Lambda=0$, for a 
triangulated manifold \cite{Freidelb,Freidelc}.  One can rewrite the action 
in terms of a set of discrete variables: a Lie algebra element $e_\alpha$ 
corresponding to the integral of $e$ along the edge $\alpha$ of a tetrahedron 
in the triangulation, and a holonomy $g_\alpha$ of the connection $\omega$ 
around the edge.  The path integral then becomes an integral over these 
variables.  As in the continuum path integral of section \ref{Lorpi}, the 
integral over the $e_\alpha$ produces a delta function $\delta(g_\alpha)$ 
for each edge.  This translates back to the geometric statement that the
constraints require the connection $\omega$ to be flat, and thus to have
trivial holonomy around a contractible curve surrounding an edge.  

For the Euclidean Ponzano-Regge action, $g\in\mathrm{SU}(2)$, and the key 
trick is now to use the Plancherel formula to express each $\delta(g_\alpha)$  
as a sum over the characters of finite-dimensional representations of 
$\mathrm{SU}(2)$.  Fairly straightforward arguments then permit an exact 
evaluation of the remaining integrals over the $g_\alpha$, reproducing the 
$6j$ symbols in the Ponzano-Regge action.  To obtain a Lorentzian version, 
one must replace $\mathrm{SU}(2)$ by $\mathrm{SO}(2,1)$.  The corresponding 
Plancherel formula involves a sum over both the (continuous) principle series 
of representations of $\mathrm{SO}(2,1)$ and the discrete series.  Consequently, 
edges may now be labeled either by discrete or continuous spins.  Similar
methods may be used for supergravity \cite{Livine}.

The resulting rather complicated expression for the partition function
may be found in \cite{Freidelb}.  The appearance of both continuous and 
discrete labels has a nice physical interpretation \cite{Freidel3}: continuous 
representations describe spacelike edges, and seem to imply a continuous length 
spectrum, while discrete representations label timelike edges, and suggest 
discrete time.  These results should probably not yet be considered conclusive, 
since they require operators that do not commute with all of the constraints, 
but they are certainly suggestive.  

While spin foam models ordinarily assume a fixed spacetime topology, recent
work has suggested a method for summing over all topologies as well, thus 
allowing quantum fluctuations of spacetime topology \cite{Freideld}.  These
results will be discussed in section \ref{Epi}.  Methods from 2+1
dimensions have also been generalized to higher dimensions, leading to
new insights into the construction of spin foams.

\subsection{Lattice methods II: Dynamical triangulations \label{DT}}

Spin foam models are based on a fixed triangulation of spacetime, with 
edge lengths serving as the basic gravitational variables.  An alternative 
scheme is ``dynamical triangulation,'' in which edge lengths are fixed 
and the path integral is represented as a sum over triangulations.  (For 
reviews of this approach in arbitrary dimensions, see \cite{Ambjorn,Lollrev}.)  
Dynamical triangulation has been proven to be quite useful in two-dimensional 
gravity, and some important steps have been taken in higher dimensions, 
especially with the recent progress in understanding Lorentzian triangulations. 

The starting point is now a simplicial complex, diffeomorphic to a manifold $M$, 
composed of an arbitrary collection of \emph{equilateral} tetrahedra, with  sides 
of length $a$.  Metric information is no longer contained in the  choice of edge 
lengths, but rather depends on the combinatorial pattern.  Such a model is not 
exact in 2+1 dimensions, but one might hope that as $a$ becomes small and the 
number of tetrahedra becomes large it may be possible to approximate an arbitrary 
geometry.  In particular, it is plausible (although not rigorously proven) that
a suitable model lies in the same universality class as genuine (2+1)-dimensional
gravity, in which case the continuum limit should be exact.

The Einstein-Hilbert action for such a theory takes the standard Regge form
(\ref{bf1}), which for spherical spatial topology reduces to a sum  
\beq
I = -k_0N_0 + k_3N_3 ,
\label{bg1}
\eeq
where $N_0$ and $N_3$ are the numbers of vertices and tetrahedra in the 
triangulation, $k_0 = a/4G$, and $k_3$ is related to the cosmological constant.
As the number of tetrahedra becomes large, the number of distinct triangulations
(the ``entropy'') increases exponentially, while the $N_3$ term in (\ref{bg1})
provides an exponential suppression.  The ``Euclidean'' path integral
$\sum\exp\{-I\}$ should thus converge for $k_3$ greater than a critical value 
$k_3^c(k_0)$.  As $k_3$ approaches $k_3^c(k_0)$ from above, expectation
values of $N_3$ will diverge, and one may hope for a finite-volume continuum 
limit as $a\rightarrow0$.

For ordinary ``Euclidean'' dynamical triangulations, few signs of such a
continuum limit have been seen.  The system appears to exhibit two phases---a
``crumpled'' phase, in which the Hausdorff dimension is extremely large,
and a ``branched polymer'' phase---neither of which look much like a classical
spacetime \cite{Lollrev}.  An alternative ``Lorentzian'' model, introduced by 
Ambj{\o}rn and Loll \cite{AmbLolla,AmbLollb,AmbLollc,AmbLolld,AmbLolle,AmbLollg},
however, has much nicer properties, including a continuum limit that appears
numerically to match a finite-sized, spherical ``semiclassical'' configuration.

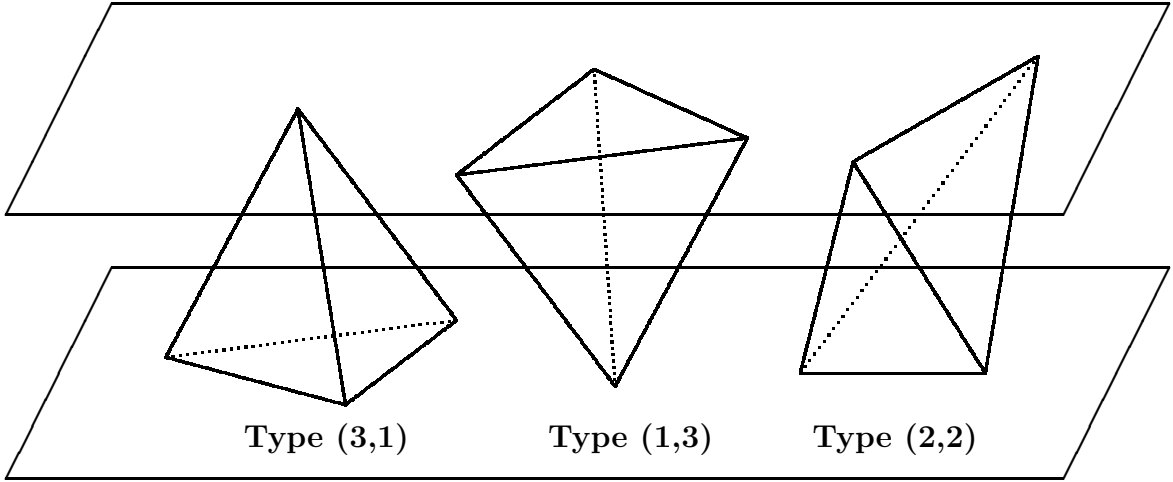
\begin{figure}
\begin{center}
\begin{picture}(400,140)(0,-80)
\thicklines
\put(-20,0){\line(1,2){40}}
\put(-20,0){\line(1,0){400}}
\put(20,80){\line(1,0){400}}
\put(380,0){\line(1,2){40}}

\put(-20,-100){\line(1,2){40}}
\put(-20,-100){\line(1,0){400}}
\put(20,-20){\line(1,0){400}}
\put(380,-100){\line(1,2){40}}

\qbezier(90,40)(65,-7)(40,-54)
\qbezier(90,40)(99,-16)(108,-72)
\qbezier(90,40)(120,0)(150,-40)
\qbezier(150,-40)(129,-56)(108,-72)
\qbezier[40](150,-40)(95,-47)(40,-54)
\qbezier(40,-54)(74,-63)(108,-72)

\qbezier(210,-65)(180,-25)(150,15)
\qbezier(210,-65)(235,-18)(260,29)
\qbezier[50](210,-65)(206,-5)(202,55)
\qbezier(150,15)(205,22)(260,29)
\qbezier(150,15)(176,35)(202,55)
\qbezier(260,29)(231,42)(202,55)

\qbezier(280,-60)(325,-60)(350,-60)
\qbezier(300,20)(335,40)(370,60)
\qbezier(280,-60)(290,-20)(300,20)
\qbezier[50](280,-60)(325,0)(370,60)
\qbezier(350,-60)(325,-20)(300,20)
\qbezier(350,-60)(360,0)(370,60)

\put(70,-88){\bf Type (3,1)}
\put(185,-88){\bf Type (1,3)}
\put(285,-88){\bf Type (2,2)}
\end{picture}
\end{center}
\caption{\small Three tetrahedra can occur in Lorentzian dynamical triangulation.
\label{tetra}}
\end{figure}

The Lorentzian model begins with a slicing of spacetime into constant time
surfaces, each of which is given an equilateral triangulation.  The region
between two neighboring slices is then filled in by tetrahedra, which can
come only in the three varieties shown in figure \ref{tetra}.  This set-up 
automatically restricts spacetime to have the topology $\mathbb{R}\times 
\Sigma$, and by declaring each slice to be spacelike and each edge joining 
adjacent slices to be timelike, one has a well-defined ``Wick rotation''
to a Riemannian signature metric with Regge action (\ref{bg1}).  Note
that for convergence, this method requires a positive value of $k_3$, and 
thus a positive cosmological constant.

The path integral for such a system can be evaluated numerically, using
Monte Carlo methods and a set of ``moves'' that systematically change an
initial triangulation \cite{AmbLollc,AmbLolld}.  One finds two phases.  At 
strong coupling, the system splits into uncorrelated two-dimensional spaces, 
each well-described by two-dimensional gravity.  At weak coupling, however, 
a ``semiclassical'' regime appears that resembles the picture obtained from 
other approaches to (2+1)-dimensional gravity.  In particular, one may evaluate 
the expectation value $\langle A(t)\rangle$ of the spatial area at fixed time 
and the correlation $\langle A(t)A(t+1)\rangle$ of successive areas; the results 
match the classical de Sitter behavior for a spacetime $\mathbb{R}\times S^2$ 
quite well.  The more ``local'' behavior---the Hausdorff dimension of a 
constant time slice, for example---is not yet well-understood.  Neither is
the role of moduli for spatial topologies more complicated than $S^2$, 
although initial steps have been taken for the torus universe \cite{Dittrich}.

The Lorentzian dynamical triangulation model can also be translated into a
two-matrix model, the so-called $ABAB$ model.  The Feynman diagrams of the
matrix model correspond to dual graphs of a triangulation, and matrix model
amplitudes become particular sums of transfer matrix elements in the gravitational
theory \cite{AmbLollf,AmbLollh,AmbLolli}.  In principle, this connection can
be used to solve the gravitational model analytically.  While this goal has
not yet been achieved (though see \cite{AmbLolli}), a number of interesting
analytical results exist.  For example, the matrix model connection can be used 
to show that Newton's constant and the cosmological constant are additively
renormalized \cite{AmbLollh}, and to analyze the apparent nonrenormalizability
of ordinary field theoretical approach.

\subsection{Other lattice approaches}

In principle the discrete approaches described in section \ref{Eda}---in
particular, the lattice descriptions of 't Hooft and Waelbroeck---should
be straightforward to quantize.  In practice, there has been fairly little
work in this area, and most of the literature that does exist involves
point particles rather than closed universes.  't Hooft has emphasized
that the Hamiltonian in his approach is an angle, and that time should
therefore be discrete \cite{tHoofte}, in agreement with the Lorentzian
spin foam analysis of section \ref{PR}.  't Hooft has also found that for 
a particular representation of the commutation relations for a point particle
in (2+1)-dimensional gravity, space may also be discrete \cite{tHooftf},
although it remains unclear whether these results can be generalized beyond 
this one special example.  Criscuolo et al.\ have examined Waelbroeck's
lattice Hamiltonian approach for the quantized torus universe \cite{Crisc},
investigating the implication of the choice of an internal time variable,
and Waelbroeck has studied the role of the mapping class group \cite{Waelq}.

\subsection{The Wheeler-DeWitt equation}

The approaches to quantization of sections \ref{reduced}--\ref{loopvar} 
share an important feature: all are ``reduced phase space'' quantizations, 
quantum theories based on the true physical degrees of freedom of the 
classical theory.  That is, the classical constraints have been solved
before quantizing, eliminating classically redundant ``gauge'' degrees 
of freedom.  In Dirac's approach to quantization \cite{Dirac,Diracb,Diracc},
in contrast, one quantizes the entire space of degrees of freedom of classical 
theory, and only then imposes the constraints.  States are initially determined
from the full classical phase space; in the ADM formulation of quantum gravity, 
for instance, they are functionals $\Psi[g_{ij}]$ of the full spatial metric.  
The constraints then act as operators on this auxiliary Hilbert space; the 
physical Hilbert space consists of those states that are annihilated by the 
constraints, with a suitable new inner product, acted on by physical operators 
that commute with the constraints.  For gravity, in particular, the Hamiltonian 
constraint acting on states leads to a functional differential equation, the 
Wheeler--DeWitt equation \cite{DeWittb,Wheeler}.

In the first order formalism, it is straightforward to show that Dirac
quantization is equivalent to the Chern-Simons quantum theory we
have already seen.  Details can be found in chapter 8 of \cite{Carlipbook},
but the basic argument is fairly clear: at least for $\Lambda=0$, the 
first order constraints coming from (\ref{ac3})--(\ref{ac4}) are at most
linear in the momenta, and are thus uncomplicated to solve.

In the second order formalism, matters become considerably more
complicated \cite{Carlip10}.  We begin with a wave function $\Psi[g_{ij}]$, 
upon which we wish to impose the constraints (\ref{ad3}), with momenta 
acting as functional derivatives,
\beq
\pi^{ij} = -i\frac{\delta\ }{\delta g_{ij}} .
\label{bi1}
\eeq
The first difficulty is that we are no longer allowed to choose a nice
time-slicing such as York time; that would be a form of gauge-fixing, and
is not permitted in Dirac quantization.  We can still decompose the
spatial metric and momentum as in (\ref{ad4}), but only up to a spatial 
diffeomorphism, which depends on an undetermined vector field $Y^i$  
appearing in the momentum $\pi^{ij}$ \cite{Mon}.  The momentum 
constraint fixes $Y^i$ in terms of the scale factor $\lambda$, but it
does so nonlocally.  As a consequence, the Hamiltonian constraint 
becomes a nonlocal functional differential equation, and very little
is understood about its solutions, even for the simplest case of the
torus universe.  Further complications come from the fact that the inner 
product on the space of solutions of the Wheeler-DeWitt equation must 
be gauge-fixed \cite{Woodard,Marolfb}; again, little is understood 
about the resulting Hilbert space.

In view of the difficulty in finding exact solutions to the Wheeler-DeWitt
equation, it is natural to look for perturbative methods, for example
an expansion in powers of Newton's constant $G$.  One can solve
the momentum constraints order by order by insisting that each term 
depend only on (nonlocal) spatially diffeomorphism-invariant
quantities.  Such an expansion has been studied by Banks, Fischler, 
and Susskind for the physically trivial topology $\mathbb{R}\times S^2$ 
\cite{Banks}, following much earlier work by Leutwyler \cite{Leut}.  Even 
in this simple case, computations quickly become extremely difficult.
Other attempts have been made to write a discrete version of the 
Wheeler-DeWitt equation in the Ponzano-Regge formalism of section \ref{PR} 
\cite{Barrettb,Makela}.  This approach has the advantage that the spatial
diffeomorphisms have already been largely eliminated, removing the main
source of nonlocality discussed above.  The Wheeler-DeWitt-like equation 
in \cite{Barrettb} has been shown to agree with the the Ponzano-Regge 
model.

\subsection{Lorentzian path integrals \label{Lorpi}}

Up to now, I have mainly concentrated on approaches to quantum
gravity that fall under the broad heading of canonical quantization.
An alternative approach---already implicit in the discussion of
discrete methods---starts with the Feynman path integral, or ``sum 
over histories.''   In an important sense, path integral methods are less 
precise than those of canonical quantization: the infinite-dimensional 
``integral'' over histories can rarely be rigorously defined, we do not
really know what classes of paths to sum over, and ordering ambiguities
in the operator formalism reemerge as ambiguities in the integration
measure.  On the other hand, path integrals allow us to ask questions%
---for example, about amplitudes for spatial topology change---that
are difficult or impossible to formulated in a canonical theory.

The simplest path integral approach to (2+1)-dimensional quantum
gravity is the phase space path integral, in which the action is written
in the ADM form (\ref{ad2})--(\ref{ad3}) and the spatial metric $g_{ij}$
and momentum $\pi^{ij}$ are treated as independent integration
variables.  The lapse and shift $N$ and $N^i$ appear as Lagrange
multipliers, and the integrals over these quantities yield delta functionals 
for the constraints $\mathcal{H}$ and $\mathcal{H}_i$.  One might
therefore expect the result to be equivalent to the canonical quantization
of section \ref{reduced}, in which the constraints are set to zero and
solved for the physical degrees of freedom.  This is indeed true, as
shown in \cite{Carlip11,Seriu} for spatially closed universes and
\cite{Cantini} for geometries with point particles.  The main subtlety
comes from the appearance of many different determinants, arising from
gauge-fixing and from the delta functionals, which must be shown
to cancel.  The phase space path integral for the first order formulation
similarly reproduces the corresponding canonically quantized theory.

It is perhaps more interesting to look at the covariant metric path 
integral, in which one starts with the ordinary Einstein-Hilbert action 
and gauge-fixes the full (2+1)-dimensional diffeomorphism group.
This approach does not require a topology $\mathbb{R}\times\Sigma$,
and could potentially describe topology-changing amplitudes.
Unfortunately, very little is yet understood about this approach.  Section 
9.2 of Ref.\ \cite{Carlipbook} describes a partial gauge-fixing, which
takes advantage of the fact that every metric on a three-manifold is
conformal to one of constant scalar curvature.  But while this leads to
some simplification, we are still left with an infinite-dimensional integral 
about which very little can yet be said.

By far the most useful results in the path integral approach to
(2+1)-dimensional quantum gravity have come from the covariant 
first-order action (\ref{ac2}).  The path integral for this action was first 
fully analyzed in two seminal papers by Witten \cite{Wittena,Wittenb},
who showed that it reduced to a ratio of determinants that has an
elegant topological interpretation as the analytic or Ray-Singer torsion 
\cite{RaySinger}.  The partition function for a closed three-manifold
with $\Lambda=0$ takes the form
\beq
Z_M  
  = \frac{|\mathop{det}\Delta_{\bar\omega}^{(3)}|_{\phantom{\bar\omega}}^{3/2}
  |\mathop{det}\Delta_{\bar\omega}^{(1)}|_{\phantom{\bar\omega}}^{1/2}}{
  |\mathop{det}\Delta_{\bar\omega}^{(2)}| }
\label{bj1}
\eeq
where $\Delta_{\bar\omega}^{(n)} = D_{\bar\omega}*D_{\bar\omega}* + 
*D_{\bar\omega}*D_{\bar\omega}$ is the gauge-covariant Laplacian acting
on $n$-forms and $\bar\omega$ is a flat $\mathrm{SO}(2,1)$ connection.
When $M$ admits more than one such flat connection, (\ref{bj1}) must be
integrated over the moduli space of such connections.  This integral sometimes
diverges \cite{Wittenb}; the significance of that divergence is not understood.

Although it was originally derived for closed manifolds, (\ref{bj1}) can be
extended to manifolds with boundary in a straightforward manner.
The path integral then gives a transition function that depends on specified
boundary data---most simply, the induced spin connection $\omega$,
with some additional restrictions on the normal component of $\omega$ 
and the triad $E$ \cite{Wu,CarCos}.  For a manifold with the topology 
$\mathbb{R}\times\Sigma$, the results agree with those of covariant
canonical quantization: the transition amplitude between two surfaces
with prescribed spin connections is nonzero only if the holonomies agree.  

But the path integral can also give transition amplitudes between states
on surfaces $\Sigma_i$ and $\Sigma_f$ with different topologies.  If we 
demand that the initial and final surfaces be nondegenerate and spacelike, 
their topologies are severely restricted: Amano and Higuchi have shown that 
$\Sigma_i$ and $\Sigma_f$ must have equal Euler numbers \cite{Amano}. For 
such manifolds, concrete computations can exploit the topological 
invariance of the Ray-Singer torsion.  Ref.\ \cite{CarCos}, for example, 
explicitly computes amplitudes for a transition between a genus three surface 
and a pair of genus two surfaces.

\subsection{Euclidean path integrals and quantum cosmology \label{Epi}}

Lorentzian path integrals allow us to compute interesting topology-changing 
amplitudes, in which the universe tunnels from one spatial topology to another.  
They do not, however, directly address a principle issue of quantum 
cosmology, the problem of describing the birth of a universe from ``nothing.''
Here, most of the literature has focused on the Hawking's Euclidean
path integral \cite{Hawking} and the Hartle-Hawking ``no boundary''
proposal \cite{Hartle}, which describes the universe in terms of a path 
integral over Riemannian metrics on manifolds with a single, connected
boundary $\Sigma$.  As in 3+1 dimensions, most of the work in 2+1
dimensions has concentrated on the saddle point approximation.  So far,
the main benefit of the lower-dimensional model has been the possibility
of treating topology more systematically, revealing interesting effects 
that are only now being explored in 3+1 dimensions.

In the Hartle-Hawking approach to quantum cosmology, the initial
wave function of the universe is described by a path integral for a compact 
manifold $M$ with a single spatial boundary $\Sigma$, as in figure~\ref{fig3}.
\begin{figure}
\begin{picture}(100,160)(-115,0)
\qbezier(21,20)(0,55)(16,100)
\qbezier(95,20)(115,56)(102,102)
\qbezier(21,20)(58,-26)(95,20)
\qbezier(16,100)(20,120)(12,140)
\qbezier(102,102)(98,121)(106,140)
\qbezier(12,140)(60,115)(106,140)
\qbezier(12,140)(60,165)(106,140)
\put(54,136){$\Sigma$}
\put(57,60){$M$}
\end{picture}
\caption{\small A manifold $M$ with a single boundary $\Sigma$ describes the
birth of a universe in the Hartle--Hawking approach to quantum cosmology.
\label{fig3}}
\end{figure}
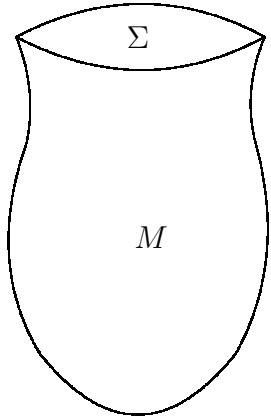
In 2+1 dimensions, the selection rules of Ref.\ \cite{Amano} imply that such 
a process can be described in Lorentzian signature only if $\chi(\Sigma)=0$,
that is, only for $\Sigma$ a torus.  Moreover, the known examples of such
metrics always yield a degenerate metric on $\Sigma$.  If one allows 
Riemannian signature, on the other hand, such a path integral makes sense
for any spatial topology, and if one further requires that $\Sigma$ be totally 
geodesic---that is, that the extrinsic curvature of $\Sigma$ vanish---one can 
smoothly join on a Lorentzian metric at $\Sigma$ \cite{Gibbons}.  Hartle and 
Hawking therefore propose a ``ground state'' wave function
\beq
\Psi[h,\varphi|_{\Sigma};\Sigma] =
  \sum_{M:\partial M=\Sigma} \int [dg][d\varphi]\, \exp\left\{-I_E[g,\varphi]\right\} ,
\label{bk1}
\eeq
where the value of the path integral is determined by a specified induced
metric $h$ and matter configuration $\varphi|_\Sigma$ on the boundary.
The summation represents a sum over topologies of $M$; in the absence of any
basis for picking out a preferred topology, all manifolds with a given boundary 
$\Sigma$ are assumed to contribute.  The wave function $\Psi$ is to be interpreted 
as an amplitude for finding a universe, with a prescribed spatial topology 
$\Sigma$, characterized by an ``initial'' geometry $h$ and a matter configuration 
$\varphi|_{\partial M}$.  This approach finesses the question of initial conditions 
for the universe by simply omitting an initial boundary, and it postpones the 
question of the nature of time in quantum gravity: information about time is hidden 
in the boundary geometry $h$, but the path integral can be formulated without 
making a choice of time explicit.

The path integral (\ref{bk1}) cannot, in general, be evaluated exactly, even in
2+1 dimensions.  Indeed, there are general reasons to expect the expression to
be ill-defined: a conformal excitation $g_{\mu\nu}\rightarrow e^{2\phi}g_{\mu\nu}$
contributes to $I_E$ with the wrong sign, and the action is unbounded below
\cite{Gibbonsb}.  In the (2+1)-dimensional Lorentzian dynamical triangulation 
models of section \ref{DT}, however, it is known that these wrong sign
contributions are unimportant\cite{AmbLollc}; they are overwhelmed by the 
much larger number of well-behaved geometries in the path integral.  This has 
led to a suggestion \cite{Dasgupta,Dasguptab} that the conformal contribution 
is canceled by a Faddeev-Popov determinant (see also \cite{Mazur}), and some 
preliminary supporting computations have been made in a proper time 
gauge \cite{Dasgupta}.

Assuming that the ``conformal factor problem'' is solved, a saddle point
evaluation of the path integral is arguably a good approximation.
For simplicity, let us ignore the matter contribution to the wave function.
Saddle points are then Einstein manifolds, with actions proportional to the 
volume.  An easy computation shows that the leading contribution to (\ref{bk1}) 
is a sum of terms of the form
\beq
\exp\left\{ -{\bar I}_E\right\} 
  = \Delta_M\exp\left\{ \mathrm{sign}(\Lambda)
  \frac{\mathit{Vol}_{\bar g}(M)}{4\pi G\hbar|\Lambda|^{1/2}}\right\} ,
\label{bk2}
\eeq
where $\bar g$ is an Einstein metric on $M$, $\mathit{Vol}_{\bar g}(M)$ is 
the volume of $M$ with the metric rescaled to constant curvature $\pm1$,
and the prefactor $\Delta_M$ is related, as in section \ref{Lorpi}, to the
Ray-Singer torsion of $M$.

For $\Lambda>0$, three-manifolds that admit Einstein metrics are all
elliptic---that is, they have constant positive curvature, and can be described
as quotients of the three-sphere by discrete groups of isometries.  The largest
value of $\mathit{Vol}_{\bar g}(M)$ comes from the three-sphere itself, and
one might expect it to dominate the sum over topologies.  As shown in
\cite{Carlip12}, though, the number of topologically distinct lens spaces 
with volumes less than $\mathit{Vol}_{\bar g}(S^3)$ grows fast enough
that these spaces dominate, leading to a divergent partition function
for closed three-manifolds.  The
implications for the Hartle-Hawking wave function have not been examined
explicitly, but it seems likely that a divergence will appear there as well.

For $\Lambda<0$, three-manifolds that admit Einstein metrics are 
hyperbolic, and the single largest contribution to (\ref{bk2}) comes
from the {\em smallest\/} such manifold.  This contribution has been worked 
out in detail, for a genus 2 boundary, in \cite{Fujiwarab}.  Here, too, 
however, manifolds with larger volumes---although individually exponentially 
suppressed---are numerous enough to lead to a divergence in the partition 
function \cite{Carlip12}.  In this case, the Hartle-Hawking wave function has 
been examined as well, and it has been shown that the wave function acquires
infinite peaks at certain specific spatial geometries: again, topologically
complicated manifolds whose individual contributions are small occur
in large enough numbers to dominate the path integral, and ``entropy''
wins out over ``action'' \cite{Carlip13}.

The benefit of restricting to 2+1 dimensions here is a bit 
different from the advantages seen earlier.  We are now helped not 
so much by the simplicity of the geometry (although this helps in
the computation of the prefactors $\Delta_M$), but by the fact that 
three-manifold topology is much better understood than four-manifold topology.  
It is only quite recently that similar results for sums over topologies 
have been found in four dimensions \cite{Carlip14,Carlip15,RatTsch1,Anderson}.

As noted in section \ref{PR}, recent work on spin foams has also suggested
a new nonperturbative approach to evaluating the sum over topologies.  Building 
on work by Boulatov \cite{Boulatov}, Freidel and Loupre have recently considered
a variant of the Ponzano-Regge model, and have shown that although the sum 
over topologies diverges, it is Borel summable \cite{Freideld}.  This result 
involves a clever representation of a spacetime triangulation as a Feynman 
graph in a field theory on a group manifold, allowing the sum over topologies 
to be reexpressed as a sum of field theory Feynman diagrams.  The model
considered in \cite{Freideld} is not exactly the Ponzano-Regge model, and
it is not clear that it is really ``ordinary'' quantum gravity.  Moreover,
study of the physical meaning of the Borel resummed partition
function has barely begun.  Nonetheless, these results suggest that a
full treatment of the sum over topologies in (2+1)-dimensional quantum 
gravity may not be hopelessly out of reach.  

There are also indications that string theory might have something
to say about the sum over topologies \cite{Farey}.  In particular, the AdS/CFT
correspondence may impose boundary conditions that limit the topologies allowed
in the sum.  Whether such results can be extended to spatially closed
manifolds remains unclear.

\section{What Have We Learned? \label{whata}}

The world is not (2+1)-dimensional, and (2+1)-dimensional quantum gravity
is certainly not a realistic model of our Universe.  Nonetheless, the
(2+1)-dimensional model reflects many of the fundamental conceptual issues 
of real world quantum gravity, and work in this field has provided some
valuable insights.\\

\noindent{\bf Existence and nonuniqueness}:\\[.5ex]
Perhaps the most important lesson of (2+1)-dimensional quantum gravity is 
that general relativity can, in fact, be quantized.  While additional 
ingredients---strings, for instance---may have their own attractions, they 
are evidently not necessary for the existence of quantum gravity.  More than 
an ``existence theorem,'' though, the (2+1)-dimensional models also provide a 
``nonuniqueness theorem'':  many approaches to the quantum theory are possible, 
and they are not all equivalent.  This is perhaps a bit of a disappointment, 
since many in this field had hoped that once we found a self-consistent 
quantum theory of gravity, the consistency conditions might be stringent enough 
to make that theory unique.  In retrospect, though, we should not be so surprised:
quantum gravity is presumably more fundamental than classical general relativity, 
and it is not so strange to learn that more than one quantum theory can 
have the same classical limit.\\

\noindent{\bf (2+1)-dimensional gravity as a test bed}:\\[.5ex]
General relativity in 2+1 dimensions has provided a valuable test bed for
a number of specific proposals for quantum gravity.  Some of these are 
``classics''---the Wheeler-DeWitt equation, for instance, and reduced phase 
space quantization---while others, like spin foams, Lorentzian dynamical 
triangulations, and covariant canonical quantization, are less well
established.  

We have discovered some rather unexpected features, such as the difficulties 
caused by spatial diffeomorphism invariance and the consequent nonlocality 
in Wheeler-DeWitt quantization, and the necessity of understanding the
representations of the group of large diffeomorphisms in almost all approaches.  
For particular quantization programs, (2+1)-dimensional models have also 
offered more specific guidance: special properties of the loop operators 
(\ref{bb1}), methods for treating noncompact groups in spin foam models, and 
properties of the sums over topologies described in section \ref{Epi} have 
all been generalized to 3+1 dimensions.\\ 

\noindent{\bf Lorentzian dynamical triangulations}:\\[.5ex]
A particular application of (2+1)-dimensional gravity as a test bed 
is important enough to deserve special mention.  The program of ``Lorentzian 
dynamical triangulations'' described in section \ref{DT} is a genuinely new 
approach to quantum gravity.  Given the failures of ordinary ``Euclidean
dynamical triangulations,'' one might normally be quite skeptical of such
a method.  But the success in reproducing semiclassical states in 2+1
dimensions, although still fairly limited, provides a strong argument that
the approach should be taken seriously.\\

\noindent{\bf Observables and the ``problem of time''}:\\[.5ex]
One of the deepest conceptual difficulties in quantum gravity has been 
the problem of reconstructing local, dynamical spacetime from the nonlocal
diffeomorphism-invariant observables required by quantum gravity.  The notorious
``problem of time'' is a special case of this more general problem of observables.
As we saw in section \ref{obs}, (2+1)-dimensional quantum gravity points toward a
solution, allowing the construction of families of ``local'' and ``time-dependent''
observables that nevertheless commute with all constraints.  

The idea that ``frozen time'' quantum gravity is a Heisenberg picture 
corresponding to a fixed-time-slicing Schr{\"o}dinger picture is a central 
insight of (2+1)-dimensional gravity.  In practice, though, we have also seen 
that the transformation between these pictures relies on our having a detailed 
description of the space of classical solutions of the field equations.  We 
cannot expect such a fortunate circumstance to carry over to full (3+1)-dimensional 
quantum gravity; it is an open question, currently under investigation, whether 
one can use a perturbative analysis of classical solutions to find suitable 
approximate observables \cite{Basu}.\\

\noindent{\bf Singularities}:\\[.5ex]
It has long been hoped that quantum gravity might smooth out the singularities of
classical general relativity.  Although the (2+1)-dimensional model has not yet
provided a definitive test of this idea, some progress has been made.  Puzio,
for example, has shown that a wave packet initially concentrated away from
the singular points in moduli space will remain nonsingular \cite{Puzio2}.  On
the other hand, Minassian has recently demonstrated \cite{Minassian} that 
quantum fluctuations do not ``push singularities off to infinity'' (as suggested 
in \cite{Hosoya}), and that several classically singular (2+1)-dimensional 
quantum spacetimes also have singular ``quantum $b$-boundaries.''\\

\noindent{\bf Is length quantized?}:\\[.5ex]
Another long-standing expectation has been that quantum gravity will lead to 
discrete, quantized lengths, with a minimum length on the order of the Planck length.  
Partial results in quantum geometry and spin foam approaches to (2+1)-dimensional 
quantum gravity suggest that this may be true, but also that the problem is a bit 
subtle \cite{Matschull,tHoofte,Colosi}.  The most recent result in this area, 
Ref.\ \cite{Freidel3}, relates the spectrum of lengths to representations of 
the (2+1)-dimensional Lorentz group, which can be discrete or continuous.  
Freidel et al.\ argue that spacelike intervals are continuous, while timelike 
intervals are discrete, with a spectrum of the form $\sqrt{n(n-1)}\ell_p$.  The 
analysis is a bit tricky, since the length ``observables'' do not, in general, 
commute with the Hamiltonian constraint.  A first step towards defining truly 
invariant operators describing distances between point particles supports this 
picture \cite{Nouib}, but the results are not yet conclusive.\\

\noindent{\bf ``Doubly special relativity''}:\\[.5ex]
Quantum gravity contains two fundamental dimensionful constants, the Planck length
$\ell_p$ and the speed of light $c$.  This has suggested to some that special
relativity might itself be altered so that both $\ell_p$ and $c$ are constants.
This requires a nonlinear deformation of the Poincar{\'e} algebra, and leads to
a set of theories collectively called ``doubly special relativity'' \cite{Amelino,%
Mag,Kowalski}.  It has recently been pointed out that (2+1)-dimensional gravity 
automatically displays such a deformation \cite{Matschull,Amelinob,FreiKow}.  A
few attempts have been made to connect this picture to noncommutative spacetime,
mainly in the context of point particles \cite{Matschull,Welling2,Balles}, 
but it seems too early to evaluate them.\\

\noindent{\bf Topology change}:\\[.5ex]
Does consistent quantum gravity require spatial topology change?  The answer in 
2+1 dimensions is unequivocally no: canonical quantization gives a perfectly 
consistent description of a universe with a fixed spatial topology.  On the other 
hand, the path integrals of section \ref{Lorpi} seem to allow the computation of 
amplitudes for tunneling from one topology to another.  Problems with 
these topology-changing amplitudes remain, particularly in the regulation
of divergent integrals over zero-modes.  If these can be resolved, 
however, we will have to conclude that we have found genuinely and deeply 
inequivalent quantum theories of gravity.\\

\noindent{\bf Sums over topologies}:\\[.5ex]
In conventional descriptions of the Hartle-Hawking wave function, and in other 
Euclidean path integral descriptions of quantum cosmology, it is usually assumed 
that a few simple contributions dominate the sum over topologies.  The results of
(2+1)-dimensional quantum gravity indicate that such claims should be treated 
with skepticism; as discussed in section \ref{Epi}, the sum over topologies is
generally dominated by an infinite number of complicated topologies, each 
individually exponentially suppressed.  This is a new and unexpected result,
whose implications for realistic (3+1)-dimensional gravity are just starting to
be explored.

\section{What Can We Still Learn? \label{whatb}}

We know immensely more about (2+1)-dimensional quantum gravity than we did
twenty years ago.  But we still have an enormous amount to learn.  In 
particular, it is only quite recently that the general tools developed over 
the past few years have been brought to bear on particular \emph{physical}
problems---the resolution of singularities, for example, and the question
of whether space is discrete at the Planck scale.  A sketchy and rather
personal list of open questions would include the following:\\

\noindent{\bf Singularities}:\\[.5ex]
A key question in quantum gravity is whether quantized spacetime ``resolves''
the singularities of classical general relativity.  This is a difficult
question---already classically, it is highly nontrivial to even \emph{define}
a singularity \cite{Clarke}, and the quantum extensions of the classical 
definitions are far from obvious.  This is an area in which (2+1)-dimensional
gravity provides a natural arena, but results so far are highly
preliminary \cite{Puzio2,Minassian}.\\

\noindent{\bf Sums over topologies}:\\[.5ex]
Another long-standing question in quantum gravity is whether spacetime topology
can (or must) undergo quantum fluctuations.  As we saw in section
\ref{Epi}, some real progress has been made in 2+1 dimensions.  Often,
though, the results require saddle point approximations, and 
pick out particular classes of saddle points.  The nonperturbative
summation techniques discussed at the end of section \ref{Epi} promise 
much deeper results, and may point toward a measure on the space of
topologies analogous to the measure on the space of geometries induced by the
DeWitt metric.\\

\noindent{\bf Quantized geometry}:\\[.5ex]
We saw above that there is some evidence for quantization of
timelike intervals in (2+1)-dimensional gravity.  A systematic exploration of this
issue might teach us a good deal about differences among approaches to quantization.
In particular, it would be very interesting to see whether any corresponding result
appears in reduced phase space quantization, Wheeler-DeWitt quantization, or path
integral approaches.  To address this problem properly, one must introduce genuine
observables for quantities such as length and area, either by adding point particles
\cite{Nouib} or by looking at shortest geodesics around noncontractible cycles.
Note that for the torus universe, the moduli can be considered as ratios of lengths,
and there is no sign that these need be discrete.  This does not contradict
the claims of \cite{Freidel3}, since the lengths in question are spacelike, but
it does suggest an interesting dilemma in Euclidean quantum gravity, where spacelike
as well as timelike intervals might naturally be quantized \cite{Rovellix}.\\

\noindent{\bf Euclidean v.\ Lorentzian gravity}:\\[.5ex]
In the Chern-Simons formalism of section \ref{first}, ``Euclidean'' and ``Lorentzian''
quantum gravity seem to be dramatically inequivalent: they have different gauge 
groups, different holonomies, and very different behaviors under the actions of
large diffeomorphisms.  In the ADM approach of section \ref{adm}, on the other hand, 
the differences are almost invisible.  This suggests that further study might  
finally tell us whether Euclideanization is merely a technical trick,
analogous to Wick rotation in ordinary quantum field theory, or whether it gives a
genuinely different theory; and, if the latter, just how different the Euclidean
and Lorentzian theories are.  In canonical quantization, a key step would be
to relate Chern-Simons and ADM amplitudes in the Euclidean theory, perhaps using the 
methods of section \ref{obs}.  In spin foam and path integral approaches, it might
be possible to explicitly compare amplitudes.\\

\noindent{\bf Which approaches are equivalent?}\\[.5ex]
A more general problem is to understand which of the approaches described  
here are equivalent.  In particular, it is not
obvious how much of the difference among various methods of quantization can be
attributed to operator ordering ambiguities, and how much reflects a deeper 
inequivalence, as reflected (for instance) in different length spectra or 
different possibilities for topology change.  An answer might help us
understand just how nonunique quantum gravity in higher dimensions will be.\\

\noindent{\bf Higher genus}:\\[.5ex]
Most of the detailed, explicit results in (2+1)-dimensional quantum gravity 
hold only for the torus universe $\mathbb{R}\times T^2$.  As noted in section 
\ref{tor}, this topology has some exceptional features, and might not be completely
representative.  In particular, the relationship between the ADM and Chern-Simons 
quantizations in section \ref{obs} relied on a particularly simple operator 
ordering; it is not obvious that such an ordering can be found for the higher
genus case \cite{Mon2}.  An extension to arbitrary genus might be too difficult, 
but a full treatment of the genus two topology, using the relation to hyperelliptic 
curves or the sigma model description of \cite{Valtancoli}, may be possible.  It 
could also be worthwhile to further explore the case of spatially nonorientable 
manifolds \cite{Loukox} to see whether any important new features arise.\\

\noindent{\bf Coupling matter}:\\[.5ex]
This review has dealt almost exclusively with vacuum quantum gravity.  We know
remarkably little about how to couple matter to this theory.  Some limited 
progress has been made: for example, there is some evidence that (2+1)-dimensional 
gravity is renormalizable in the $1/N$ expansion when coupled to scalar fields  
\cite{Kugo,Mizoguchic}.  This is apparently no longer the case when gravity is
coupled to fermions and a $\mathrm{U}(1)$ Chern-Simons gauge theory \cite{Anselmi}, 
although Anselmi has argued that if coupling constants are tuned to exact values,  
renormalizability can be restored, and in fact the theory can be made finite
\cite{Anselmib}.  Certain matter couplings in supergravity have been studied 
\cite{deWit,Matschullb}, and work on circularly symmetric ``midi-superspace 
models'' has led to some surprising results, including unexpected bounds on
the Hamiltonian \cite{Ashtekare,AshtekarPierri,Gambinib,Beetle,Varadarajanc,Pierri}.  
But the general problem of coupling matter remains very difficult, not least 
because---except in the special case of ``topological matter'' \cite{Geg1,Geg2}---we 
lose the ability to represent diffeomorphisms as $\mathrm{ISO}(2,1)$ gauge 
transformations.  

Difficult as it is, however, an understanding of matter couplings may be 
the key to many of the conceptual issues of quantum gravity.  One can explore 
the properties of a singularity, for example, by investigating the reaction 
of nearby matter, and one can look for quantization of time by examining 
the behavior of physical clocks.  Moreover, some of the deep questions of 
quantum gravity can be answered only in the presence of matter.  For example,
does gravity cut off ultraviolet divergences in quantum field theory? 
This idea is an old one \cite{DeWitt,Ishamc,Ishamd}, and it gets
some support from the boundedness of the Hamiltonian in midi-superspace
models \cite{AshtekarPierri}, but it is only in the context of a full
quantum field theory that a final answer can be given.\\

\noindent{\bf The cosmological constant}:\\[.5ex]
Undoubtedly, the biggest embarrassment in quantum gravity today is the
apparent prediction, at least in effective field theory, that the cosmological
constant should be some $120$ orders of magnitude larger than the observed 
limit.  Several attempts have been made to address this problem in the context 
of (2+1)-dimensional quantum gravity.  First, Witten has suggested a novel
mechanism by which supersymmetry in 2+1 dimensions might cancel radiative
corrections to $\Lambda$ without requiring the equality of superpartner
masses, essentially because even if the vacuum is supersymmetric, the 
asymptotics forbid the existence of unbroken supercharges for massive
states \cite{Wittene,StromBecker}.  This argument requires special features
of 2+1 dimensions, though, and it is not at all clear that it can be generalized
to 3+1 dimensions (although some attempts have been made in the context of
``deconstruction'' \cite{Jejjala}).  

Second, the discovery that the sum over topologies can lead to a divergent 
partition function has been extended to 3+1 dimensions, at least for $\Lambda<0$, 
and it has been argued that this behavior might signal a phase transition that 
could prohibit a conventional cosmology with a negative cosmological constant 
\cite{Carlip14,Carlip15}.  The crucial case of a positive cosmological 
constant is not yet understood, however, and if a phase change does indeed 
occur, its nature is still highly obscure.  It may be that the nonperturbative 
summation over topologies discussed at the end of section \ref{Epi} could cast 
light on this question.  

One might also hope that a careful analysis of the coupling of matter in 2+1 
dimensions could reveal useful details concerning the vacuum energy contribution 
to $\Lambda$, perhaps in a setting that goes beyond the usual effective field 
theory approach.  For example, there is evidence that the matter Hamiltonian 
is bounded above in (2+1)-dimensional gravity \cite{Ashtekare}; perhaps this 
could cut off radiative contributions to the cosmological constant at an 
interesting scale.\\

\noindent{\bf Again, (2+1)-dimensional gravity as a test bed}:\\[.5ex]
As new approaches to quantum gravity are developed, the (2+1)-dimensional model
will undoubtedly remain important as a simplified test bed.  For example, a bit
of work has been done on the null surface formulation of classical gravity in
2+1 dimensions \cite{Forni}; a quantum treatment might be possible, and could
tell us more about the utility of this approach in 3+1 dimensions.  Similarly,
(2+1)-dimensional gravity has recently been examined as an arena in which to 
test for a new partially discrete, constraint-free formulation of quantum 
gravity \cite{Gambinic}.

\vspace{1.5ex}
\begin{flushleft}
\large\bf Acknowledgments
\end{flushleft}

I would like to thank Giovanni Arcioni, John Barrow, Roman Jackiw,
Nemanja Kaloper, Jorma Louko, and Vincent Moncrief for useful comments.
This work was supported in part by Department of Energy grant
DE-FG02-91ER40674.

\end{document}